\documentclass[%
 reprint,
nofootinbib,
 amsmath,amssymb,
 aps,
]{revtex4-2}
\usepackage{adjustbox}
\usepackage{xcolor}
\usepackage{mathtools}
\usepackage{braket}
\usepackage{soul}
\usepackage{amsmath}
\usepackage{multirow}
\usepackage{relsize}
\usepackage{tabularx}
\usepackage{array}
\usepackage{soul}
\usepackage{graphicx}
\usepackage{dcolumn}
\usepackage{bm}
\usepackage{hyperref}
\usepackage{ upgreek }

\usepackage{tikz}

\begin{document}

\preprint{APS/123-QED}

\title{Stochastic modeling of superfluorescence in compact systems}

\author{Stasis Chuchurka$^\dagger$}%
\email{stasis.chuchurka@desy.de}
\author{Vladislav Sukharnikov}%
\thanks{These authors contributed equally to this work.}
\affiliation{
Deutsches Elektronen-Synchrotron DESY, 22603 Hamburg, Germany\;
\\
Department of Physics, Universität Hamburg, 22761 Hamburg, Germany
}%

\author{Andrei Benediktovitch}
\affiliation{
Deutsches Elektronen-Synchrotron DESY, 22603 Hamburg, Germany
}%

\author{Nina Rohringer}
\email{nina.rohringer@desy.de}
\affiliation{
Deutsches Elektronen-Synchrotron DESY, 22603 Hamburg, Germany\,
\\
Department of Physics, Universität Hamburg, 22761 Hamburg, Germany
}%

\date{\today}


\begin{abstract}
    We propose an approach based on stochastic differential equations to describe superfluorescence in compact ensembles of multi-level emitters in the presence of various incoherent processes. This approach has a numerical complexity that does not depend on the number of emitters. The stochastic differential equations are derived directly from the quantum master equation. In this study, we present a series of numerical examples, comparing our solution to exact calculations and discussing the limits of applicability. For many relevant cases, the proposed stochastic differential equations provide accurate results and correctly capture quantum many-body correlation effects.
\end{abstract}


\maketitle

\section{\label{sec: introduction} Introduction}

Superfluorescence is a notable phenomenon in quantum optics, that is observed when incoherently excited atoms collectively emit radiation in the form of a highly energetic and short burst of light. The initially produced spontaneous emission couples the dipole moments of atoms, allowing them to synchronize the emission of photons. The phenomenon of superfluorescence traces its origins back to the seminal work of Dicke~\cite{1954'Dicke}. Since then, it has garnered significant theoretical interest~\cite{PhysRevA.2.2038, PhysRevA.2.883, Gross1982, Bullough1987, PhysRevA.81.053821, PhysRevResearch.5.013091} and has been a source of inspiration for numerous experimental studies. These studies include early demonstrations in gases~\cite{PhysRevLett.30.309, PhysRevLett.39.547, PhysRevLett.42.224, Cahuzac1979} and solids~\cite{PhysRevLett.59.1189}, as well as more recent demonstrations in quantum dots~\cite{Rain2018}, nitrogen-vacancy centers~\cite{Bradac2017}, cold atoms~\cite{PhysRevLett.116.083601, PRXQuantum.3.010338}, and nuclei~\cite{doi:10.1126/science.1187770}. Earlier investigations spanned from optical~\cite{Cahuzac1979} and infrared~\cite{Flusberg1976, Gross1976} to millimeter~\cite{Gross1979, Moi1983} wavelengths. The recent emergence of X-ray free-electron lasers has provided exciting opportunities to observe the phenomenon of superfluorescence in the X-ray domain~\cite{Rohringer2012, PhysRevLett.111.233902, 2015'Yoneda, PhysRevLett.120.133203, PhysRevLett.123.023201}. Nevertheless, these new opportunities bring additional complications to the theoretical description of the underlying phenomena. For example, X-ray transitions often undergo significant decoherence in the form of the Auger-Meitner effect, which can intensely compete with the excitation process and substantially disrupt the synchronization of atomic dipoles. Therefore, the theoretical models that previously proved efficient need thorough revision.

In Ref.~\cite{1954'Dicke}, Dicke introduced a minimal model necessary for observing collective spontaneous emission. This model comprises a system of identical two-level quantum emitters interacting with a quantized electromagnetic field assumed to be uniform across the ensemble. Being indistinguishable, $N$ emitters evolve through a ladder of $(N + 1)$ collective many-body states, synchronizing the radiation phases of the different emitters. This basic model can be generalized to include multi-level emitters~\cite{Silva2022} and incoherent processes~\cite{Sukharnikov2023, gegg2016}, which only partially diminish the collective nature of the interaction with the field. Providing invaluable insight from a theoretical point of view, these approaches, however, do not suggest an efficient strategy for numerical studies. Decomposing the quantum states in the basis set that accounts for the permutation symmetry results in a system of equations, the number of which grows polynomially with the number of atoms $N$. Although the resulting polynomial complexity allows treating a moderately large number of atoms ($N\lesssim 100$) with a relatively acceptable computational effort~\cite{Sukharnikov2023}, the problems involving a realistically large number of atoms still remain challenging.

This numerical complexity has been acknowledged previously, as experiments often involve macroscopic numbers of atoms $N\gg 1$. As outlined in Ref.~\cite{Gross1982}, when the system is instantaneously excited, and there are no competing incoherent processes, quantum effects dominate in the early stages of superfluorescence. Subsequently, the evolution becomes classical, and we can effectively simulate the dynamics of quantum emitters by solving Bloch equations with statistically distributed initial dipole moments. By considering multiple regions with independently distributed initial conditions, we can conduct a numerical analysis of macroscopic distributed systems.

However, if the initial incoherent excitation triggering superfluorescence is not instantaneous, and the quantum stage is further complicated by various incoherent processes, random initial conditions are no longer applicable. In such cases, several phenomenological strategies have been proposed. For instance, in Refs.~\cite{2004'Ziolkowski,2000'Larroche,2019'Subotnik-EhrenfestR,2019'Subotnik_comparison}, random initial conditions have been replaced by phenomenological noise terms acting as source terms in the Maxwell-Bloch equations. Nevertheless, as these methods are not derived from first principles, they come with certain limitations. For example, the widely used methodology proposed in Ref.~\cite{2000'Larroche} produces an incorrect temporal profile of spontaneous emission, as highlighted in Refs.~\cite{2018'Krusic, 2019'Benediktovitch}.

In summary, there is currently no numerically efficient and sufficiently accurate methodology available for providing a reliable quantitative characterization of collective spontaneous emission involving an arbitrary number of multi-level emitters, especially in the presence of incoherent processes and excitation. Therefore, our objective is to establish such a formalism based on first principles. The development of this formalism draws inspiration from a many-body phase-space description utilizing the positive $P$ function. (for more details and examples, see Refs.~\cite{2014'Drummond_book, 2001CoPhC.142..442D, PhysRevE.96.013309, PhysRevLett.98.120402, PRXQuantum.2.010319}). {While this approach offers a direct path to stochastic equations that can be efficiently sampled in a Monte Carlo style, unfortunately, it features certain limitations that manifest in practice as spiky, diverging solutions. These limitations and a potential method for mitigating them are thoroughly discussed in Refs.~\cite{PhysRevA.55.3014, 2006'Deuar_stochastic-gauges, 2005'Deuar_PhD}. In these works, it has been demonstrated that quantum many-body systems allow certain freedom when modeled by stochastic differential equations. It turns out that there is more than one system of equations leading to the same expectation values. In practice, different systems of equations may exhibit different degree of divergent behaviour. The technique, which provides several strategies for choosing a more stable system of equations, is commonly referred to as stochastic gauges. The present article offers an analysis of how the mentioned instability issue impacts the simulation of superfluorescence and how we adopt stochastic gauges to resolve them.
}

Given the need to benchmark the proposed theoretical methodology, this article focuses exclusively on superfluorescence in compact systems. This choice is motivated by the fact that it can be exactly solved using methods based on the decomposition of the quantum state. Specifically, we adopt the methodology presented in Ref.~\cite{Sukharnikov2023}. The numerical benchmark suggests that the proposed methodology provides satisfactory results for a wide range of parameter values. Discrepancies, however, arise when the system evolves into a dark many-body state. Furthermore, we demonstrate that the presence of incoherent processes mitigates the prominence of this issue.

In our formalism, neither the form nor the number of equations depends on the number of emitters $N$ in the system. $N$ only enters the equations as a parameter, making the methodology free from the numerical difficulties inherent in techniques based on quantum state decomposition. For the extended methodology suitable for analyzing distributed systems, we refer the interested reader to Ref.~\cite{chuchurka2023stochastic}.

Let us outline the structure of the article. In Sec.~\ref{sec: mean field approximation}, we construct the master equation for superfluorescence in compact systems. In Sec.~\ref{sec: Stochastic equations}, we rephrase the quantum-mechanical problem in terms of stochastic differential equations. We also address numerical challenges encountered during the simulations and propose potential solutions. In Sec.~\ref{sec: Numerical analysis}, we analyze various conditions under which the phenomenon of superfluorescence can be observed. We begin with the simplest example in Sec.~\ref{Cooperative emission of two-level atoms}, involving the cooperative emission of instantly excited two-level atoms. By comparing our simulations with those based on the methodology presented in Ref.~\cite{Sukharnikov2023}, we evaluate the performance of our proposed method for various numbers of atoms and initial conditions. We demonstrate that under specific circumstances when the system evolves into a dark many-body state, our methodology fails to reproduce the correct behavior. When the influence of such states is not significant, we achieve adequate results with minimal computational resources. In Sec.~\ref{Sec: quasi lambda}, we explicitly include excitation via incoherent pumping. This accentuates the challenge posed by dark states. We demonstrate how this issue can be mitigated by introducing decoherence typically present in experimental conditions. In Secs. \ref{Quantum beats in V system} and \ref{Lasing in Lambda system}, we conclude the numerical examples by studying multi-level effects in superfluorescence observed in $V$- and $\Lambda$-systems. In this case, we demonstrate the performance of the methodology by constructing non-trivial three-particle correlation functions. In Sec. \ref{discussion}, we give an overview of the accuracy, and efficiency of our methodology and share empirical observations made during the numerical studies.

\section{\label{sec: mean field approximation} Compact systems}

The dynamics of the field in distributed systems is generally complex and depends on many factors that are insignificant for our goal to introduce the key ideas of our stochastic formalism applied to superfluorescence. In Ref.~\cite{1954'Dicke}, it was supposed that in a compact ensemble of atoms, the system size was considerably smaller than the wavelength of the field. As a result, the atoms saw an identical field. This simplified model neglects the influence of dipole-dipole interactions between the atoms that was shown to be detrimental to observing superfluorescence \cite{Friedberg1974, Gross1982, 2003'Clemens, 2016'Sutherland}. In certain cases of strong dipole-dipole interactions, the so-called dipole blockade \cite{Comparat2010}, suppresses the electronic transitions initiated by a narrow-band field. The effects of dipole-dipole interactions strongly depend on the geometry and distances between the emitters, thus defining the minimal interatomic distance beyond which the superradiant behavior takes place \cite{2022'Masson, 2022'Sierra}.

Numerical analysis with fully implemented dipole-dipole interactions is quite involved since it requires individual treatment of each atom due to the broken symmetry. One possible workaround is to replace the dipole-dipole interacting atoms with atoms that interact only through the radiative field but possess different transition frequencies, mimicking the impact of the dipole-dipole interactions \cite{Andreoli2021}. Another approach is to reduce the effect of the dipole-dipole interactions by introducing a bad cavity or an elongated dilute atomic system. The former approach will ultimately lead us to superfluorescence in compact systems. As shown in Ref. \cite{Goldstein1997}, proper use of a cavity eliminates the effect of dipole-dipole interactions, which may explain a good agreement between experiments in Refs. \cite{Raimond1982, Laske2019} and a simple Dicke model without any account of dipole-dipole interactions. Indeed, a cavity selects optical wave vectors close to the transition frequency $\omega_0$, which filters out the dipole-dipole interactions and simplifies the spatial dependence of the field. Based on these assumptions, the problem of superfluorescence is considerably simplified and reduces to a Dicke master equation that we aim to solve by means of stochastic differential equations.

Consider a system of $N$ identical multi-level atoms characterized by a system of levels $\{\ket{p}\}$, energies $\hbar\omega_p$, and the following free Hamiltonian:
\begin{equation*}
    \hat{H}_0=\sum_p\hbar\omega_p\sum_a \hat{\sigma}_{a,pp}.
\end{equation*}
Here, we utilize the operators $\hat{\sigma}_{a,pq}=\ket{p}_a\!\bra{q}_a$ to describe transitions between states for each atom $a$. Initially, atoms are uncorrelated and described by a density matrix symmetric under any permutation.

The atomic levels are coupled to the quantized electric displacement field $\hat{\textbf{D}}\!\left(\textbf{r}\right)$. The presence of the cavity makes its amplitude uniform across the sample. We suppose a field of single carrier wave vector $\textbf{k}_0$ associated with the transition frequency $\omega_0=ck_0$ given by
\begin{multline*}
    \hat{\textbf{D}}(\textbf{r})\approx \hat{\textbf{D}}^{\!(+)}\!e^{i\textbf{k}_0\textbf{r}}+\hat{\textbf{D}}^{\!(-)}\!e^{-i\textbf{k}_0\textbf{r}} \\= \sum_\lambda \left(D_0 \hat{a}_{\lambda}\textbf{e}_{\lambda}e^{i\textbf{k}_0\textbf{r}}+D_0^*\hat{a}^\dag_{\lambda}\textbf{e}_{\lambda}^* \,e^{-i\textbf{k}_0\textbf{r}}\right).
\end{multline*}
Here, $D_0=i\sqrt{{\hbar\omega_0\varepsilon_0}/{[2V]}}$, $V$ is the quantization volume, $\hat{a}_{\lambda}$ and $\hat{a}^\dagger_{\lambda}$ are the bosonic field operators, and the vectors $\textbf{e}_{\lambda}$ are the polarizations of the field perpendicular to $\textbf{k}_0$. Among all the atomic levels, the light only couples two subsets: the ground state manifold ${\ket{g}}$ and the excited state manifold ${\ket{e}}$. Their energy splittings, denoted as $\omega_{ee'}=\omega_e-\omega_{e'}$ and $\omega_{gg'}=\omega_g-\omega_{g'}$, are assumed to be much smaller than the carrier frequency $\omega_0$. Later in this article, we use indices $p, q, r, s,i,j$ to represent any arbitrary state, while specifically reserving indices $g$ and $e$ for states from the ground and excited state manifolds, respectively. The dynamics of the atomic populations is supposed to change on time scales large compared to $1/\omega_0$, so that non-resonant contributions are neglected. Based on these approximations, we write the following interaction Hamiltonian
\begin{equation}
\label{eq: expression for the V operator}
    \begin{multlined}
        \hat{V}=-\frac{1}{\varepsilon_0}\hat{\textbf{D}}^{\!(+)}\sum_{e,g}\textbf{d}_{eg}\sum_a\hat{\sigma}_{a,eg}e^{i\textbf{k}_0\textbf{r}_a}+\text{h.c.}
  \end{multlined}
\end{equation}
where $\textbf{d}_{pq}$ are the matrix elements of the dipole moment operators. The atomic coherences assemble in the sum $\sum_a\hat{\sigma}_{a,eg}e^{i\textbf{k}_0\textbf{r}_a}$ reflecting the collective interaction with the field mode\footnote{In Ref. \cite{Raimond1982}, the atomic sample is positioned at an antinode of a standing wave, which can be taken into account by neglecting $e^{\pm i\textbf{k}_0\textbf{r}_a}$.}. Henceforth, we omit the multiplier $e^{i\textbf{k}_0\textbf{r}_a}$ since it can be adjusted by shifting the phases of the states ${\ket{e}}$ and ${\ket{g}}$. We introduce collective dipole moments $\hat{\textbf{P}}^{(\pm)}$ composed of the phased operators $\hat{\sigma}_{a,pq}$ as follows:
\begin{equation*}
    \begin{aligned}
        \hat{\textbf{P}}^{(-)}&=\sum_{eg}\textbf{d}_{eg}\sum_a\hat{\sigma}_{a,eg},
        \\
        \hat{\textbf{P}}^{(+)}&=\sum_{eg}\textbf{d}_{ge}\sum_a\hat{\sigma}_{a,ge}.
    \end{aligned}
\end{equation*}
This gives a compact expression for the interaction Hamiltonian in Eq. (\ref{eq: expression for the V operator}):
\begin{equation}
\label{eq: expression for the V operator 2}
    \hat{V}=-\frac{1}{\varepsilon_0}\left(\hat{\textbf{P}}^{\!(-)}\hat{\textbf{D}}^{\!(+)}+\hat{\textbf{P}}^{\!(+)}\hat{\textbf{D}}^{\!(-)}\right).
\end{equation}

Having only one mode in a cavity simplifies the spatial dynamics; however, it can only lead to optical phenomena such as quantum Rabi oscillations, collapse, and revivals \cite{scully1999quantum, Eberly1980}, which we do not intend to analyze in this article. Additionally, achieving collective spontaneous emission requires a substantial leakage of photons, as discussed in, for example, Ref. \cite{Bullough1987}.

After an atom emits a photon, it makes effectively $Q/[k_0L]$ passes\footnote{$Q/[k_0L]$ is the decay rate of the intensity. The field amplitudes decay two times slower.} through the cavity before it gets damped. Here, $Q$ is the quality factor, and $L$ is the length of the cavity. In order to have collective spontaneous emission, the dynamics of the atomic populations and coherences must be much slower than the leakage of the field. In these circumstances, we can trace out the field degrees of freedom by applying the Born-Markov approximation, which leads to the well-known Dicke master equation formulated for multi-level atoms
\begin{multline}
\label{eq: Dicke master equation}
    \dfrac{d \hat{\rho}(t)}{dt}
    =\mathcal{L}[\hat{\rho}(t)]=\frac{i}{\hbar}\left[\hat{\rho}(t),\hat{H}_0+\hat{V}_{\text{in}}(t)\right]\\+\mathcal{L}_{\text{coll.}}[\hat{\rho}(t)]+\mathcal{L}_{\text{incoh.}}[\hat{\rho}(t)].
\end{multline}
Initially, the field is coupled to the atoms through the interaction Hamiltonian $\hat{V}$. After tracing out the field degrees of freedom, its role is taken over by two operators $\mathcal{L}_{\text{coll.}}[\hat{\rho}(t)]$ and $\hat{V}_{\text{in}}(t)$. The superoperator $\mathcal{L}_{\text{coll.}}[\hat{\rho}(t)]$ represents collective dissipation caused by the interaction of the atoms with \textit{their own light} from previous passes:
\begin{multline*}
    \mathcal{L}_{\text{coll.}}[\hat{\rho}(t)]=\frac{\gamma}{2}\sum_{\alpha}\bigg(\left[\hat{P}^{(+)}_\alpha\hat{\rho}\!\left(t\right),\hat{P}^{(-)}_\alpha\right]\\+\left[\hat{P}^{(+)}_\alpha,\hat{\rho}\!\left(t\right)\hat{P}^{(-)}_\alpha\right]\bigg),
\end{multline*}
where $\gamma=2Q/[V\hbar \varepsilon_0]$, and $\hat{P}^{(\pm)}_\alpha$ are components of the vector operators $\hat{\textbf{P}}^{(\pm)}$. Since the atoms interact with the light collectively, $\mathcal{L}_{\text{coll.}}[\hat{\rho}(t)]$ eventually involves only the collective dipole moments $\hat{\textbf{P}}^{(\pm)}$. 

If the system is exposed to an \textit{externally applied field} $\bm{\mathcal{D}}_{\text{in}}(\textbf{r},t)$, which is uniform across the atomic ensemble:
\begin{equation}
    \label{eq: incoming field}
    \bm{\mathcal{D}}_{\text{in}}(\textbf{r},t)=\bm{\mathcal{D}}_{\text{in}}^{(+)}(t)e^{i\textbf{k}_0\textbf{r}}+\bm{\mathcal{D}}_{\text{in}}^{(-)}(t)e^{-i\textbf{k}_0\textbf{r}},
\end{equation}
it can be included in the master equation with the interaction Hamiltonian $\hat{V}_{\text{in}}(t)$ constructed similar to $\hat{V}(t)$ in Eq. (\ref{eq: expression for the V operator 2})
\begin{equation*}
    \hat{V}_{\text{in}}(t)=-\frac{1}{\varepsilon_0}\left(\hat{\textbf{P}}^{\!(-)}\bm{\mathcal{D}}_{\text{in}}^{(+)}(t)+\hat{\textbf{P}}^{\!(+)}\bm{\mathcal{D}}_{\text{in}}^{(-)}(t)\right)
\end{equation*}
The amplitudes $\bm{\mathcal{D}}_{\text{in}}^{(\pm)}(t)$ can be deterministic complex-valued functions representing classical fields or quantum light in a coherent state. Moreover, $\bm{\mathcal{D}}_{\text{in}}^{(\pm)}(t)$ can have some statistical distribution that reproduces moments of normal-ordered field operators. The amplitudes $\bm{\mathcal{D}}_{\text{in}}^{(\pm)}(t)$ allow the inclusion of black-body photons or any arbitrary external field causing stimulated emission.

Besides collective interaction with the radiation, atoms undergo a wide variety of incoherent processes, such as non-radiative decay, ionization. By introducing a separate independent reservoir for each atom and assuming that atoms interact with them identically, we apply the Markovian approximation \cite{Meystre2007, gegg2017identical} and derive the most general form of $\mathcal{L}_{\text{incoh.}}[\hat{\rho}(t)]$:
\begin{multline}
\label{eq: incoherent part of the master equation}
    \mathcal{L}_{\text{incoh.}}[\hat{\rho}(t)]=\frac{1}{2} \;  \mathlarger{\sum}_{\mathclap{\substack{\,a,p,q,\\r,s}}} \; \Gamma_{pqrs}\!\left(t\right)\Big(\big[\hat{\sigma}_{a,pq}\hat{\rho}\!\left(t\right),\hat{\sigma}_{a,sr}\big]\\+\big[\hat{\sigma}_{a,pq},\hat{\rho}\!\left(t\right)\hat{\sigma}_{a,sr}\big]\Big).
\end{multline}
The characteristics of the incoherent processes enter the equations through the rates $\Gamma_{pqrs}\!\left(t\right)$.

\section{Stochastic equations}\label{sec: Stochastic equations}

The master equation (\ref{eq: Dicke master equation}) is symmetric under atomic permutations. When the system starts from any symmetric density matrix, the problem can be solved for a moderately large number of atoms ($N\lesssim 100$) by applying methods from Refs. \cite{Sukharnikov2023, gegg2016}. A simplified method from Ref. \cite{Silva2022} can be used when the atoms start from a statistical mixture of symmetric pure states and interact only collectively, namely, without $\mathcal{L}_{\text{incoh.}}[\hat{\rho}(t)]$. The main drawback of these methods is their polynomial scaling with $N$.

The development of our formalism draws inspiration from the concept of a positive $P$ function~\cite{2014'Drummond_book, 2001CoPhC.142..442D, PhysRevE.96.013309, PhysRevLett.98.120402, PRXQuantum.2.010319} used to generate stochastic differential equations for the problems involving bosonic fields. The final equations possess an intuitive form: the deterministic parts remind classical equations, whereas the quantum effects are attributed to the noise terms. 

We attempt to derive similar equations for a compact system of quantum emitters. First, we analyze the distinctions between the quantum description based on the master equation and the semi-classical one based on the Bloch equations~\cite{Gross1982, scully1999quantum}. Further, we demonstrate how these equations can be enhanced with supplementary stochastic terms, characterized by specific statistical properties, to restore the missing quantum properties.

\subsection{Optical Bloch equations}

\begin{figure*}[t!]
    \centering
    \includegraphics[width = 0.85\linewidth]{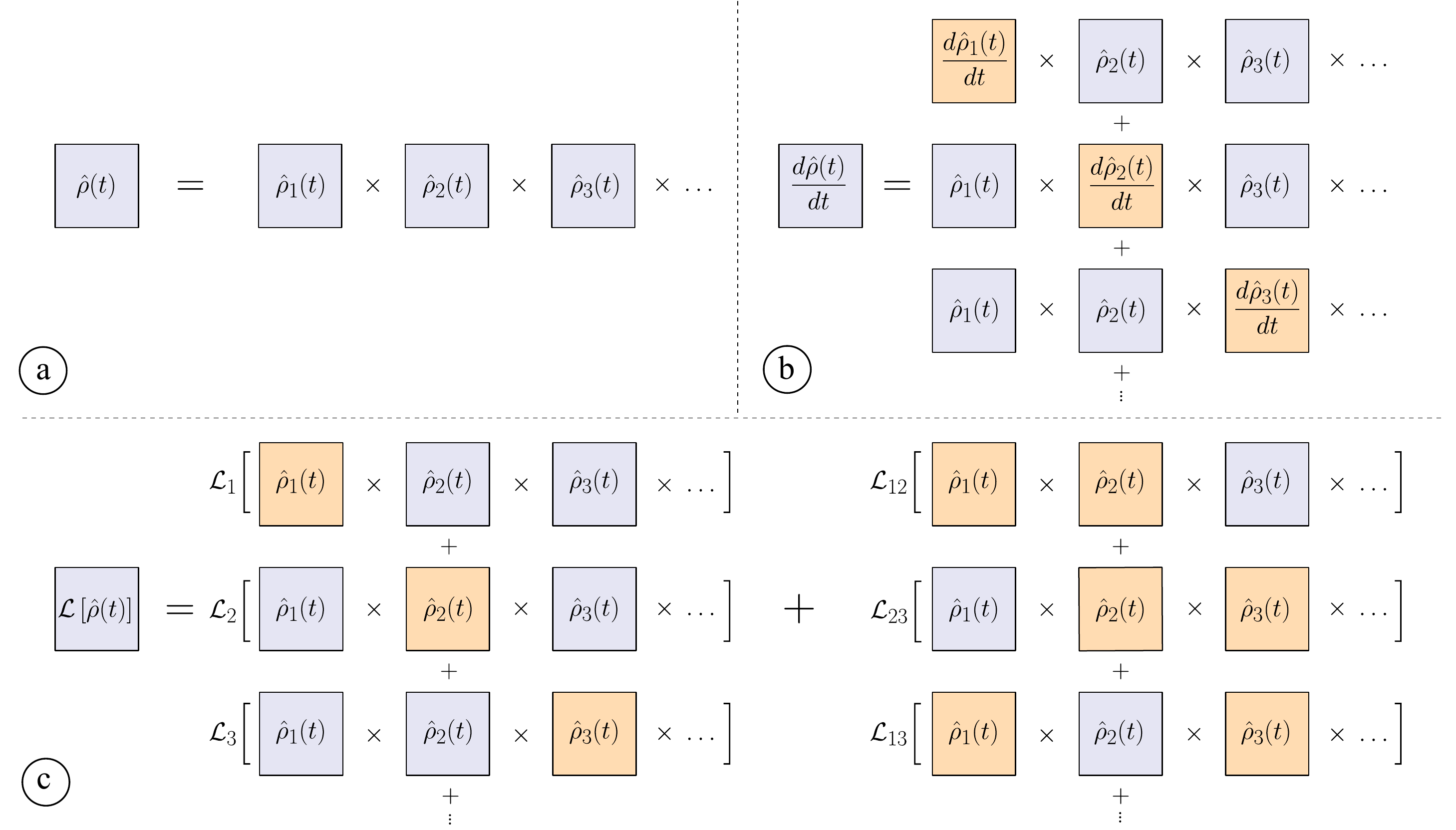}\newpage
    \caption{Schematic representation of the completely factorized density matrix (a) and its interaction with the time derivative (b) and $\mathcal{L}[\hat{\rho}(t)]$ (c). In panel (a), we illustrate the density matrix of the whole ensemble as a product state composed of individual single-particle density matrices. The time derivative of this density matrix leads to a sum of products, where only single-particle density matrices are affected, as highlighted by the orange color in panel (b). Upon applying the collective Liouville operator, in addition to products where only a single particle is affected $\mathcal{L}_{a}[\ldots]$, there is an additional contribution from two-particle interactions $\mathcal{L}_{a\not = b}[\ldots]$. This is represented in panel (c).\newline~\newline}
    \label{fig: visualization}
\end{figure*}

We start the derivation with the simplest possible ansatz for the density matrix $\hat{\rho}(t)$, assuming a complete factorization of the atomic degrees of freedom in terms of single-particle density matrices $\hat{\rho}_a(t)$:
\begin{equation}
\label{eq: classical sampling}
\hat{\rho}(t)=\prod_a \hat{\rho}_a(t).
\end{equation}
Furthermore, since the system is symmetric under permutations, we assume that all $\hat{\rho}_a(t)$ have the same matrix elements $\rho_{pq}(t)$, so that 
\begin{equation*}
\hat{\rho}_a(t) = \sum_{pq}\rho_{pq}(t)\hat{\sigma}_{a,pq}.
\end{equation*}
Fig. \ref{fig: visualization} (a) schematically depicts this ansatz. Although this decomposition is suitable for the assumed initial state of a fully symmetric density matrix of uncorrelated atoms, the further evolution of the system can only be partially captured by this proposed ansatz. To get the equations for the variables ${\rho}_{pq}(t)$, we generate the following equations for the expectation values
$\text{Tr}(\hat{\sigma}_{a,pq}\hat{\rho}(t))$:
\begin{equation*}
    \frac{d}{dt}\text{Tr}(\hat{\sigma}_{a,pq} \hat{\rho}(t))=\text{Tr}(\hat{\sigma}_{a,pq}\mathcal{L}[\hat{\rho}(t)]).
\end{equation*}
Assuming the decomposition in Eq.~(\ref{eq: classical sampling}), the expectation values $\text{Tr}(\hat{\sigma}_{a,pq}\hat{\rho}(t))$ are equal to $\rho_{qp}(t)$. The considered ansatz for the density matrix factorizes second-order correlators, namely:
\begin{equation}
\label{eq: facorization two-particle}
    \text{Tr}(\hat{\sigma}_{a,pq} \hat{\sigma}_{b,rs} \hat{\rho}(t))=\rho_{qp}(t)\rho_{sr}(t),
\end{equation}
which leads to a closed system of equations known as Bloch equations~\cite{Gross1982, scully1999quantum}:
\begin{widetext}

\begin{align}
\label{eq: bloch equations}
 \dot{\rho}_{pq}\!\left(t\right) \nonumber&=-i\omega_{pq}{\rho}_{pq}\!\left(t\right)+\frac{1}{2}\sum_{i,j} \color{black}\Big(2\Gamma_{piqj}\!\left(t\right)\rho_{ij}\!\left(t\right)-\Gamma_{ijip}\!\left(t\right)\rho_{jq}\!\left(t\right)-\rho_{pj}\!\left(t\right)\Gamma_{iqij}\!\left(t\right)\Big)\\&+\frac{\gamma}{2}\sum_{r,s}\Big(2{\rho}_{rs}(t)\textbf{d}_{p<r}\textbf{d}_{s>q}-\textbf{d}_{p>r}\textbf{d}_{r<s}{\rho}_{sq}(t)-{\rho}_{pr}(t)\textbf{d}_{r>s}\textbf{d}_{s<q}\Big)\\\nonumber &+\frac{i}{\hbar\varepsilon_0}\bm{\mathcal{D}}^{(+)}\!\left(t\right)\sum_{r}\Big(\textbf{d}_{p>r}{\rho}_{rq}\!\left(t\right)-{\rho}_{pr}\!\left(t\right)\textbf{d}_{r>q}\Big)+\frac{i}{\hbar\varepsilon_0}\bm{\mathcal{D}}^{(-)}\!\left(t\right)\sum_{r}\Big(\textbf{d}_{p<r}{\rho}_{rq}\!\left(t\right)-{\rho}_{pr}\!\left(t\right)\textbf{d}_{r<q}\Big),
\end{align}
where $p>q$ means that index $p$ corresponds to the subset of excited states $\{\ket{e}\}$ and index $q$ represents the subset of ground states $\{\ket{g}\}$. Each atom interacts with the field amplitudes $\bm{\mathcal{D}}^{(\pm)}(t)$ that combine the incoming fields $\bm{\mathcal{D}}^{(\pm)}_{\text{in}}(t)$ and the field produced by the other $N-1$ atoms:
\begin{equation}
\label{eq: fields}
    \bm{\mathcal{D}}^{(+)}(t)=\bm{\mathcal{D}}^{(+)}_{\text{in}}(t)+ i\hbar\varepsilon_0\frac{\gamma}{2} (N-1)\sum_{e,g}\textbf{d}_{ge}{\rho}_{eg}(t),\quad 
    \bm{\mathcal{D}}^{(-)}(t)=\bm{\mathcal{D}}^{(-)}_{\text{in}}(t)- i\hbar\varepsilon_0\frac{\gamma}{2} (N-1)\sum_{g,e}\textbf{d}_{eg}{\rho}_{ge}(t).
\end{equation}

The factorization of the second-order correlators in Eq.~(\ref{eq: facorization two-particle}) used in the derivations of the Bloch equations shows that these equations are valid only for systems with strong classical behavior. Let us reconstruct the neglected terms in the master equation (\ref{eq: Dicke master equation}) and analyze their structure. If we insert the decomposition from Eq. (\ref{eq: classical sampling}) in the master equation (\ref{eq: Dicke master equation}), and then apply Bloch equations (\ref{eq: bloch equations}), we notice that the right-hand side of $\mathcal{L}[\hat{\rho}(t)]$ in Eq. (\ref{eq: Dicke master equation}) is restored only partially
\begin{equation}
\label{eq: missing terms}
    \mathcal{L}[\hat{\rho}(t)]-\dfrac{d \hat{\rho}(t)}{dt}=\sum_{b\neq c}\hat{\chi}_{b,c}(t)\prod_{a\neq b,c}\hat{\rho}_a(t).
\end{equation} 
This is schematically depicted in Figs. \ref{fig: visualization} (b) and (c). The time derivative of Eq. (\ref{eq: classical sampling}) can generate the terms, where only one $\hat{\rho}_a$ is modified, as illustrated on panel (b). Consequently, the remaining terms in Eq. (\ref{eq: missing terms}) entangle pairs of $\hat{\rho}_a$ through $\hat{\chi}_{b,c}(t)$ defined as follows:
\begin{equation}
\label{eq: chi operator}
    \hat{\chi}_{b,c}(t)=\sum_{p,q,r,s}\chi_{pqrs}(t)\hat{\sigma}_{b,pq}\hat{\sigma}_{c,rs},
\end{equation}
where
\begin{equation}
\label{eq: chi}
\begin{aligned}
    \chi_{pqrs}(t)&=\frac{\gamma}{2}\bigg[\bigg(\sum_{r'}\rho_{pr'}(t)\textbf{d}_{r'>q}-{\rho}_{pq}(t)\sum_{g,e}\textbf{d}_{eg}{\rho}_{ge}(t)\bigg)\sum_{p'}\Big(\textbf{d}_{r<p'}{\rho}_{p's}(t)-{\rho}_{rp'}(t)\textbf{d}_{p'<s}\Big)\\&+ \sum_{r'}\Big({\rho}_{pr'}(t)\textbf{d}_{r'>q}-\textbf{d}_{p>r'}{\rho}_{r'q}(t)\Big)\bigg(\sum_{p'}\textbf{d}_{r<p'}{\rho}_{p's}(t)-{\rho}_{rs}(t)\sum_{e,g}\textbf{d}_{ge}{\rho}_{eg}(t)\bigg)\bigg] + [p,q \rightleftarrows r,s].
\end{aligned}
\end{equation}
{Here, the second term is generated by exchanging the pairs of indices $(p,q)$ and $(r,s)$. The structure of $\hat{\chi}_{b,c}(t)$ is schematically illustrated in Fig. 1 (c) by a second term, which explicitly shows two-particle interactions.}
\end{widetext}

\subsection{Stochastic terms}

Although $\chi_{pqrs}(t)$ looks complicated, the uncompensated terms in the right-hand side of Eq.~(\ref{eq: missing terms}) contain only entangled pairs of atoms, as Eq.~(\ref{eq: chi operator}) suggests. Non-trivial correlations of higher orders are not involved, so the terms in Eq.~(\ref{eq: chi operator}) can be correctly recaptured by adding appropriate stochastic terms to the Bloch equations (\ref{eq: bloch equations}). In addition to the deterministic time evolution, we introduce stochastic terms $F_{pq}$ as described by the following equation:
\begin{equation}
\label{eq: additional terms to the Bloch equations}
\dot{\rho}_{pq}\left(t\right)\big|_{\text{noise}}=F_{pq}(\{\rho_{ij}(t)\},t).
\end{equation}
It is important to note that the properties of the noise terms can be generally parametrized by the dynamic variables $\rho_{ij}(t)$. We only constrain $F_{pq}$ to be Gaussian white noise terms with zero mean and the following second-order correlation properties:
\begin{multline}
\label{eq: correlator def}
\big\langle F_{pq}(\{x_{ij}\},t)F_{rs}(\{x_{ij}\},t')\big\rangle \\
 = \kappa_{pqrs}(\{x_{ij}\},t)\delta(t-t').
\end{multline}
The coefficients $\kappa_{pqrs}$ will be specified later. Note that Eq.~(\ref{eq: correlator def}) is parameterized by {free} parameters $x_{ij}$, which {can take on the role of the} dynamic variables $\rho_{ij}(t)$, like, for instance, in Eq.~(\ref{eq: additional terms to the Bloch equations}). When included in the noise terms, the dynamic variables $\rho_{ij}(t)$ contribute their own statistics. To capture the statistical properties inherent solely in the noise terms $F_{pq}$, Eq.(\ref{eq: correlator def}) is formulated without explicit dependence on $\rho_{ij}(t)$.



We assume that the noise terms are integrated in Itô's sense. Typically, the stochastic equations are solved with the Monte Carlo approach. The proper statistics of the dynamic variables $\rho_{pq}\!\left(t\right)$ is reconstructed by repeatedly solving equations with a randomly sampled stochastic contribution in accordance with their statistical properties. Since the variables are independently integrated for each repetition, the problem is parallelizeable, which gives a great advantage in performance compared to the methods based on the direct decomposition of the quantum state in some basis set. 

To reconstruct the density matrix $\hat{\rho}(t)$, we insert each realization of the variables $\rho_{pq}\left(t\right)$ into the decomposition in Eq. (\ref{eq: classical sampling}) and aggregate all realizations of density matrices (\ref{eq: classical sampling}) into a normalized linear combination:
\begin{multline}
\label{eq: stochastic density matrix}
    \hat{\rho}(t)=\Big\langle\prod_a\hat{\rho}_a(t)\Big\rangle=\Big\langle\prod_a \sum_{pq}\rho_{qp}(t)\hat{\sigma}_{a,pq}\Big\rangle\\
 =\sum_{i}\left[\prod_a \sum_{pq}\rho^{(i)}_{qp}(t)\hat{\sigma}_{a,pq}\right] \big/ N_{\text{sample}}.
\end{multline}
This density matrix is no longer factorizable. Here, $\rho^{(i)}_{qp}(t)$ represents the $i$-th realization of the variables $\rho_{qp}(t)$, and $N_{\text{sample}}$ is the total number of statistical realizations. We anticipate that this linear combination can restore missing entangled terms in Eq. (\ref{eq: missing terms}). The remaining step is to identify a specific expression for $\kappa_{pqrs}(t)$.

Although the new decomposition in Eq. (\ref{eq: stochastic density matrix}) does not change the expression for the first term $\mathcal{L}[\hat{\rho}(t)]$, the derivative $\dfrac{d \hat{\rho}(t)}{dt}$ is modified by additional terms proportional to $\kappa_{pqrs}$, due to Itô's lemma. Consider an arbitrary function $S$ that depends on variables ${\rho}_{pq}(t)$. If the variables ${\rho}_{pq}(t)$ are governed by equations that include the noise terms from Eq.~(\ref{eq: additional terms to the Bloch equations}), Itô's lemma can be expressed as:
\begin{equation}
\label{app: ito}
    \frac{dS}{dt}=\sum_{p,q}\frac{\partial S}{\partial {\rho}_{pq}} \frac{d {\rho}_{pq}}{d t}+\frac{1}{2}\sum_{p,q,r,s}\frac{\partial^2 S}{\partial {\rho}_{pq}\partial{\rho}_{rs}} \kappa_{pqrs}.
\end{equation}
Consequently, the full derivative of the density matrix in Eq. (\ref{eq: stochastic density matrix}) acquires the following additional contribution:
\begin{multline*}
    \dfrac{d \hat{\rho}(t)}{dt}=...\\+\bigg\langle\sum_{b\neq c}\sum_{p,q,r,s}\kappa_{pqrs}(\{\rho_{ij}(t)\},t)\hat{\sigma}_{b,pq}\hat{\sigma}_{c,rs}\prod_{a\neq b,c}\hat{\rho}_a(t)\bigg\rangle,
\end{multline*} 
that entangles pairs of atoms and has exactly the same form as the right-hand side of Eq. (\ref{eq: missing terms}). Consequently, if the correlators of the noise terms taken as
\begin{equation*}
    \kappa_{pqrs} = \chi_{pqrs},
\end{equation*}
the Bloch equations (\ref{eq: bloch equations}), supplemented by the noise terms from Eq. (\ref{eq: additional terms to the Bloch equations}), fully satisfy the master equation (\ref{eq: Dicke master equation}). 

To simulate $F_{pq}$ numerically, we have to decompose them in terms of independent noise terms. There is no unique decomposition, however, the structure of Eq. (\ref{eq: chi}) suggests the most compact one, given by the following expression:
\begin{widetext}
    \begin{equation}
    \begin{aligned}
    \label{eq: noise decomposition}
        F_{pq}(\{\rho_{ij}(t)\}, t)&=\sqrt{\frac{\gamma}{2}}\sum_{r}\bigg[\Big(\textbf{d}_{p<r}{\rho}_{rq}(t)-{\rho}_{pr}(t)\textbf{d}_{r<q}\Big)\textbf{f}(t)+\Big({\rho}_{pr}(t)\textbf{d}_{r>q}-\textbf{d}_{p>r}{\rho}_{rq}(t)\Big)\textbf{g}(t)\bigg]\\&+ \sqrt{\frac{\gamma}{2}}\bigg(\sum_{r}{\rho}_{pr}(t)\textbf{d}_{r>q}-{\rho}_{pq}(t)\sum_{g,e}\textbf{d}_{eg}{\rho}_{ge}(t)\bigg)\textbf{f}^\dag(t)+\sqrt{\frac{\gamma}{2}}\bigg(\sum_{r}\textbf{d}_{p<r}{\rho}_{rq}(t)-{\rho}_{pq}(t)\sum_{e,g}\textbf{d}_{ge}{\rho}_{eg}(t)\bigg) \textbf{g}^{\dag}(t),
    \end{aligned}
    \end{equation}
\end{widetext}
where we introduce vectors $\textbf{f}(t)$, $\textbf{f}^\dag(t)$, $\textbf{g}(t)$, $\textbf{g}^\dag(t)$ whose components are Gaussian white noise terms independent of the dynamic variables ${\rho}_{rq}(t)$. The vectors $\textbf{f}(t)$, $\textbf{f}^\dag(t)$ are statistically independent from the vectors $\textbf{g}(t)$, $\textbf{g}^\dag(t)$. The vectors $\textbf{f}(t)$, $\textbf{f}^\dag(t)$ have the correlation properties
\begin{equation}
\label{eq: elementary noise terms}
    \begin{gathered}
        \langle f_\alpha(t)f_\beta(t')\rangle=\langle f_\alpha^\dag(t)f_\beta^\dag(t')\rangle=0,\\
        \langle f_\alpha(t)f_\beta^\dag(t')\rangle = \delta_{\alpha\beta}\delta(t-t'),
    \end{gathered}
\end{equation}
that can only be sampled by complex-valued Gaussian white noise terms. Corresponding stochastic properties hold for $\textbf{g}(t)$ and $\textbf{g}^\dag(t)$. The number of the components of the vectors $\textbf{f}(t)$, $\textbf{f}^\dag(t)$, $\textbf{g}(t)$, $\textbf{g}^\dag(t)$ is defined by the dimensionality of the dipole moment vector $\textbf{d}_{eg}$. Eq. (\ref{eq: elementary noise terms}) does not uniquely define the form of the noise terms. One can simply choose $\textbf{f}^\dag(t)$, $\textbf{g}^\dag(t)$ to be complex conjugates of $\textbf{f}(t)$, $\textbf{g}(t)$. Another freedom is given by rescaling of the noise terms: if $\textbf{f}(t)$ is divided and $\textbf{f}^\dag(t)$ is multiplied by the same number, the statistical properties in Eq. (\ref{eq: elementary noise terms}) are preserved. This freedom of choice is equivalent to the diffusion gauge in the context of positive $P$ representation formalism \cite{2005'Deuar_PhD}. The way we fix the form of the elementary noise terms $\textbf{f}(t)$, $\textbf{f}^\dag(t)$, $\textbf{g}(t)$, $\textbf{g}^\dag(t)$ is discussed in Sec. \ref{sec: stochastic gauges}. 

Adding the noise terms enriches the Bloch equations with the spontaneous nature of the quantum mechanics, allowing a correct treatment of the spontaneous emission --- an indispensable triggering process of superfluorescence. As mentioned in the introduction, many authors recognized the importance of adding stochastic terms into semi-classical equations by different phenomenological approaches~\cite{2004'Ziolkowski,2000'Larroche,2019'Subotnik-EhrenfestR,2019'Subotnik_comparison}. Our approach is based on rigorous derivation and hence can serve as a base for further investigations and approximate methods.

Note that the noise decomposition in Eq.\,(\ref{eq: noise decomposition}) conserves the ``trace'' of the effective density matrix $\rho_{pq}(t)$
\begin{equation*}
    \sum_p \dot{\rho}_{pp}(t)= \sum_p F_{pp}({\{\rho_{ij}(t)\}}, t)=0.
\end{equation*}
This property is important for generating compact expressions for the expectation values. Based on the decomposition in Eq. (\ref{eq: stochastic density matrix}), one can show that the one and two-particle expectation values possess intuitive expressions in terms of the stochastic variables $\rho_{pq}(t)$:
\begin{equation}
\label{eq: expectation values}
\begin{gathered}
    \text{Tr}(\hat{\sigma}_{a,pq}\hat{\rho}(t))= \langle\rho_{qp}(t)\rangle,
    \\ 
    \text{Tr}(\hat{\sigma}_{a,pq}\hat{\sigma}_{b,rs}\hat{\rho}(t))=\langle\rho_{qp}(t)\rho_{sr}(t)\rangle,
\end{gathered}
\end{equation}
where $a\not=b$. Similar expressions hold for high-order correlation functions.

\subsection{Stochastic freedom\label{sec: stochastic gauges} }

Unfortunately, the stochastic terms $F_{pq}(t)$ break an important property of the variables $\rho_{pq}(t)$ expected from the original, deterministic Bloch equations. Starting from Hermitian initial conditions, the Bloch equations preserve the Hermiticity of the variables $\rho_{pq}(t)$. By Hermiticity, we henceforth refer to the condition where $\rho_{pq}(t) = \rho_{qp}^*(t)$. However, in order to sample the correlator $\chi_{pqrs}(t)$, the noise terms must be non-Hermitian, that is, $F_{pq}(t) \neq F_{qp}^*(t)$, which makes the dynamic variables $\rho_{pq}(t)$ non-Hermitian as well, namely $\rho^*_{pq}(t) \neq \rho_{qp}(t)$. 

{In a broader context, the non-Hermiticity of the effective density matrix signifies a doubling of the number of independent dynamic variables compared to the anticipated semi-classical scenario, where atoms are characterized by Hermitian one-particle density matrices. This doubling of dynamic variables is also inherent in phase-space methods based on positive $P$ representation \cite{2014'Drummond_book}. }

Breaking of the Hermiticity comes with the drawback of diverging behavior of the solutions of the Bloch equations. Even without any noise terms, the original Bloch equations written for non-Hermitian variables may lead to unstable solutions with hyperbolic divergence:
\begin{equation*}
    \rho_{pq}(t) \sim \frac{1}{t - t_0}.
\end{equation*}
As the singularity is approached, the dynamic variables become anti-Hermitian, that is, $\rho_{eg}(t) = -\rho_{ge}^*(t)$. Consequently, attempting to simulate Eq. (\ref{eq: bloch equations}) with the noise terms in Eq. (\ref{eq: noise decomposition}) leads to an unstable temporal dependence of expectation values.

{ In the context of the positive $P$ representation, the same stability issues are encountered, which motivated the development of so-called stochastic gauges \cite{2006'Deuar_stochastic-gauges,2005'Deuar_PhD} (in Appendix \ref{app: stochastic gauges}, we adopt these stochastic gauges for our formalism). 
In the provided references, it has been discovered that a quantum many-body system can be modeled by more than one system of stochastic differential equations. Consequently, the preferable choice is to opt for the system of equations that demonstrates less divergent behavior, which is the key idea behind stochastic gauges. Specifically, we employ two techniques known as drift and diffusion gauges, as introduced in Refs. \cite{2006'Deuar_stochastic-gauges,2005'Deuar_PhD}.

The stochastic drift gauges allow us to alter the deterministic components of the stochastic differential equations. This modification must be compensated by an appropriate re-weighting of the stochastic trajectories. According to the drift gauging procedure reproduced in Appendix \ref{app: drift gauge}, we include a weight coefficient $\Omega(t)=e^{C_0(t)}$ in the decomposition of the density matrix in Eq. (\ref{eq: stochastic density matrix}):
}
\begin{equation}
\label{eq: weighted stochastic density matrix}
\hat{\rho}(t)=\Big\langle \Omega(t)\prod_a \sum_{pq}\rho_{qp}(t)\hat{\sigma}_{a,pq}\Big\rangle.
\end{equation}
This change is also reflected in the expressions for the expectation values in Eq. (\ref{eq: expectation values}):
\begin{equation}
\label{eq: weighted expectation values}
\begin{gathered}
    \text{Tr}(\hat{\sigma}_{a,pq} \hat{\rho}(t)) =  \langle \Omega(t) \rho_{qp}(t) \rangle,
    \\ 
    \text{Tr}(\hat{\sigma}_{a,pq}\hat{\sigma}_{b,rs}\hat{\rho}(t)) = \langle \Omega(t) \rho_{qp}(t) \rho_{sr}(t) \rangle,
\end{gathered}
\end{equation}
where $a\not=b$. 

The form of the equation for the weight coefficient directly depends on how we modify the deterministic parts. The modification of the deterministic part of the Bloch equations should {counteract} the unbounded growth of the dynamic variables $\rho_{pq}(t)$. 

{Even for the determinstic Bloch equations, one can expect divergent solutions for a small violation of Hermitcity of the dynamic variables. Consequently, when the full stochastic Bloch equations are considered, the noise terms seed this non-Hermiticity, which then leads to divergence due to the structure of the deterministic terms.

The structure of the deterministic terms can be slightly modified to ensure that it does not lead to any instability. Specifically, we implement the following substitution in Eq. (\ref{eq: fields}):}
\begin{equation}
\label{eq: field modification}
\begin{aligned}
    \sum_{g,e}\textbf{d}_{eg}{\rho}_{ge}(t) \quad \to \quad \frac{1}{2}\sum_{g,e}\textbf{d}_{eg}\left({\rho}_{ge}(t)+{\rho}_{eg}^*(t)\right),\\
    \sum_{e,g}\textbf{d}_{ge}{\rho}_{eg}(t) \quad \to \quad \frac{1}{2}\sum_{e,g}\textbf{d}_{ge}\left({\rho}_{eg}(t)+{\rho}_{ge}^*(t)\right).
\end{aligned}
\end{equation}
The fields become Hermitian after this substitution, namely $\bm{\mathcal{D}}^{(+)}(t)=\bm{\mathcal{D}}^{(-)*}(t)$. {This modification of the deterministic parts can be achieved using stochastic drift gauges, which requires introducing a weight coefficient $\Omega(t) = e^{C_0(t)}$.} According to the expressions given in Appendix \ref{app: drift gauge}, the coefficient $C_0(t)$ starts from zero and satisfies the following equation:
\begin{multline}\label{eq: weight equation}
    \frac{dC_0(t)}{dt}=\frac{N-1}{2}\sqrt{\frac{\gamma}{2}}\sum_{g,e}\Big[\textbf{f}^\dag(t)\textbf{d}_{eg}\left({\rho}_{ge}(t)-{\rho}_{eg}^*(t)\right)\\+\textbf{g}^\dag(t)\textbf{d}_{ge}\left({\rho}_{eg}(t)-{\rho}_{ge}^*(t)\right)\Big].
\end{multline}
Note that the right-hand side of this equation is proportional to the anti-Hermitian parts of the variables $\rho_{pq}(t)$. Since the weight coefficient $\Omega(t)$ involves the exponentiation of $C_0(t)$, which is itself proportional to $(N-1)$, $\Omega(t)$ can rapidly grow over time. Consequently, the averaging in Eq.~(\ref{eq: weighted expectation values}) may require a large number of statistical realizations to converge. To reduce the need for the proposed drift gauge, we introduce two additional techniques.

First, we notice that the drift gauge is not always required since the original equations do not always increase the anti-Hermitian parts of the dynamic variables. We can apply the drift gauge once the relative increase of the anti-Hermitian parts per time step exceeds a certain limit. In practice, we have found that an individual stochastic trajectory requires gauging only when the population inversions are not negative, i.e., $\text{Re}\left[\rho_{ee}(t) - \rho_{gg}(t)\right] \ge 0$ for any excited state $\ket{e}$ and ground state $\ket{g}$. Since the variables are complex at the level of single trajectories, we take the real parts of the populations.

In the context of superfluorescence, the positive sign of the population inversions causes exponential amplification of the field components $\mathcal{D}^{(\pm)}_\alpha(t)$, whereas negative population inversions lead to their absorption. Consequently, positive population inversions increase both the Hermitian and anti-Hermitian parts of the field components, which can trigger diverging behavior. Negative population inversions, in contrast, reduce the fields and their anti-Hermitian parts, making gauging unnecessary.

A second technique is based on the flexibility provided by the correlation properties of the elementary noise vectors $\textbf{f}(t)$, $\textbf{f}^\dag(t)$, $\textbf{g}(t)$, and $\textbf{g}^\dag(t)$. {This method is known as the diffusion gauge, as discussed in Refs. \cite{2006'Deuar_stochastic-gauges,2005'Deuar_PhD}. }In Eq.~(\ref{eq: elementary noise terms}), we have only outlined their correlation properties without prescribing any specific form. As mentioned before, there is no unique way to define them. One of the possible representations, which we later employ in the numerical simulations, takes the following form:
\begin{equation}
\label{eq: specific form of the elementary noise terms}
\begin{gathered}
    f_\alpha(t)=\eta_\alpha(t)\bar{f}_\alpha(t), \quad f_\alpha^\dag(t)=\eta_\alpha^{-1}(t)\bar{f}_\alpha^*(t),\\
     g_\alpha(t)=\theta_\alpha(t) \bar{g}_\alpha(t), \quad g^\dag(t)=\theta_\alpha^{-1}(t)g^*_\alpha(t),
    \end{gathered}
\end{equation}
where $\theta_\alpha(t)$ and $\eta_\alpha(t)$ can take on any values. The only constraint is that they must be statistically independent of the noise terms from the future. Eq.~(\ref{eq: specific form of the elementary noise terms}) explicitly associates the noise terms $\textbf{f}(t)$ and $\textbf{f}^\dag(t)$ with a single vector of independent complex noise terms $\bar{\textbf{f}}(t)$. Similarly, $\textbf{g}(t)$ and $\textbf{g}^\dag(t)$ are linked to $\bar{\textbf{g}}(t)$. These new noise terms, $\bar{\textbf{f}}(t)$ and $\bar{\textbf{g}}(t)$, consist of independent and normal Gaussian white real noise terms $\bar{\textbf{f}}_1(t)$, $\bar{\textbf{f}}_2(t)$, $\bar{\textbf{g}}_1(t)$, and $\bar{\textbf{g}}_2(t)$:
\begin{equation}
\label{eq: decomposition of the new elementary noise terms}
\begin{gathered}
       \bar{\textbf{f}}(t) = \frac{1}{\sqrt{2}}\left(\bar{\textbf{f}}_1(t)+i\bar{\textbf{f}}_2(t)\right),\\
       \bar{\textbf{g}}(t) = \frac{1}{\sqrt{2}}\left(\bar{\textbf{g}}_1(t)+i\bar{\textbf{g}}_2(t)\right).
\end{gathered}
\end{equation}
The explicit representation in Eq.~(\ref{eq: specific form of the elementary noise terms}) preserves the correlation properties in Eq.~(\ref{eq: elementary noise terms}) regardless of the form of $\theta_\alpha(t)$ and $\eta_\alpha(t)$. To reduce the need for the drift gauge presented in Eq.~(\ref{eq: field modification}), we fix the form of the functions $\theta_\alpha(t)$ and $\eta_\alpha(t)$ in such a way that the anti-Hermitian parts of the dipole moments
\begin{equation}
\label{eq: function to be minimized}
    \left|\sum_{g,e}d_{eg,\alpha}\left({\rho}_{ge}(t)-{\rho}_{eg}^*(t)\right)\right|
\end{equation}
are minimized for each component $\alpha$. The resulting expressions for $\theta_\alpha(t)$ and $\eta_\alpha(t)$ can be found in Appendix~\ref{app:Diffusion gauge}.

{However, in certain cases, these two techniques aimed at reducing the need for the stochastic drift gauge are insufficient to control the growth of the weight coefficient $\Omega(t)$. In these circumstances, large absolute values of the weight coefficient $\Omega(t)$ lead to spikes in temporal profiles of expectation values.  While the proposed gauging techniques do not entirely resolve the instability issue, they significantly mitigate it. To improve convergence, trajectories with a weight coefficient exceeding $e^5$ in absolute value are removed in the numerical examples given in Sec. \ref{sec: Numerical analysis}.

Besides the structure of the deterministic terms, another source of divergence exists that cannot be efficiently addressed with the stochastic drift gauge. In practice, we have observed that quadratic contributions in the noise terms, as given in Eq. (\ref{eq: noise decomposition}), can, in certain cases, cause unbounded growth of the density matrix $\rho_{pq}(t)$. When the absolute value of one of the density matrix elements exceeds 100, such realizations are removed from the statistical sample in the numerical examples presented in Sec. \ref{sec: Numerical analysis}.

The comparison with the full quantum-mechanical simulations presented in Sec. \ref{sec: Numerical analysis} shows that omitting the unstable stochastic trajectories after applying the stochastic gauges does not significantly compromise the accuracy. As demonstrated in Sec. \ref{discussion}, this strategy performs noticeably better than using the ungauged original equations~\eqref{eq: bloch equations} with the noise terms given in Eq.~\eqref{eq: noise decomposition}.}

 \begin{figure*}[ht!]
    \centering
    \includegraphics[width = \linewidth]{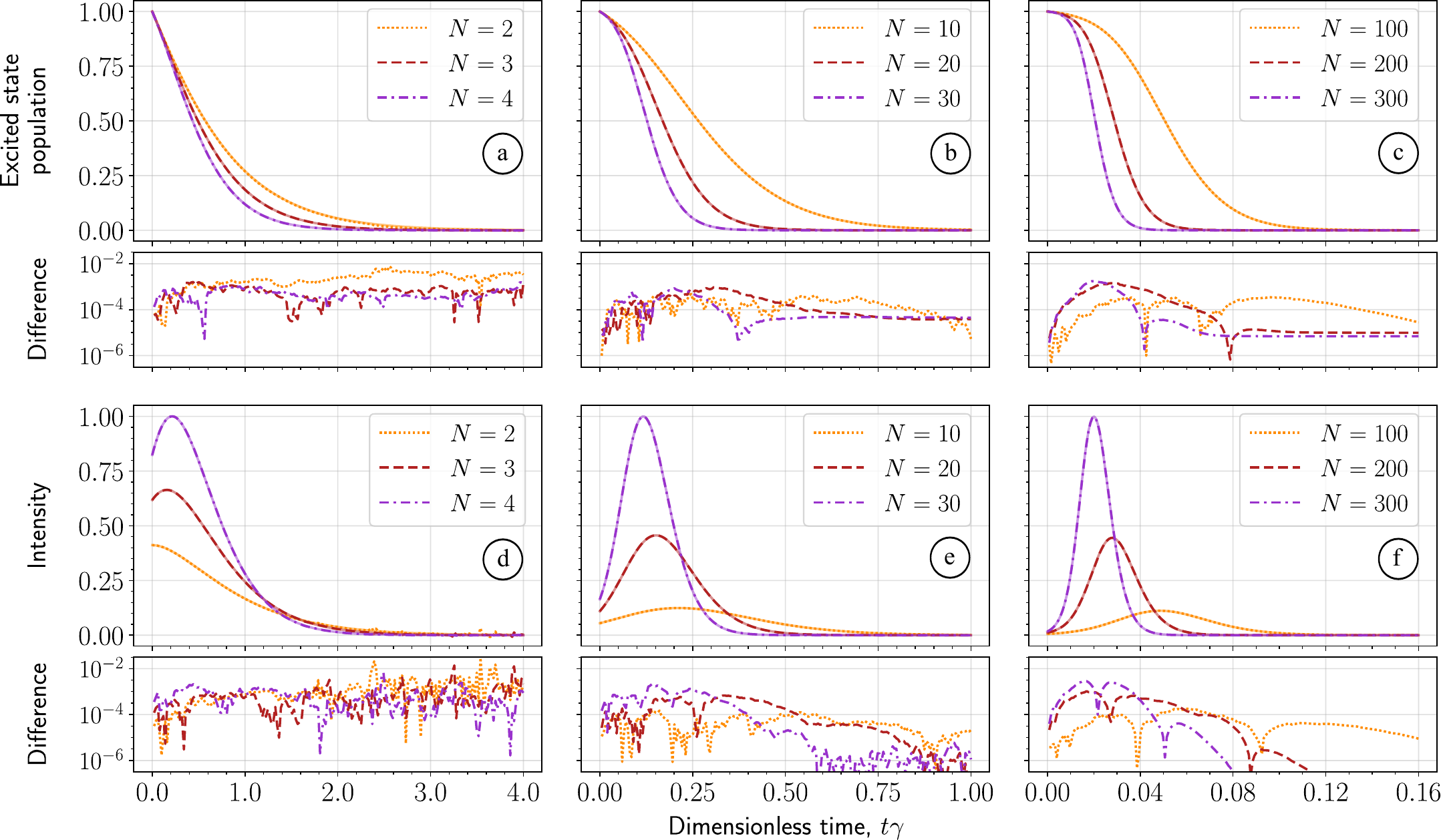}
    \caption{Solutions to the stochastic equations, modified as described in Sec. \ref{sec: stochastic gauges}. The semi-transparent lines correspond to quantum expectation values, while the opaque lines represent stochastic averages. The intensities (d, e, f) have been normalized to the maximum value in the panel. The subplots below each row show the absolute difference {between populations and normalised intensities based on} stochastic averages and quantum expectation values. For the cases of $N = 2, 3, 4$ (a, d), we have omitted { 22, 4, 3} unstable trajectories, respectively.\newline}
    \label{fig:comparison}
\end{figure*}

 \begin{figure*}[t!]
    \centering
    \includegraphics[width = \linewidth]{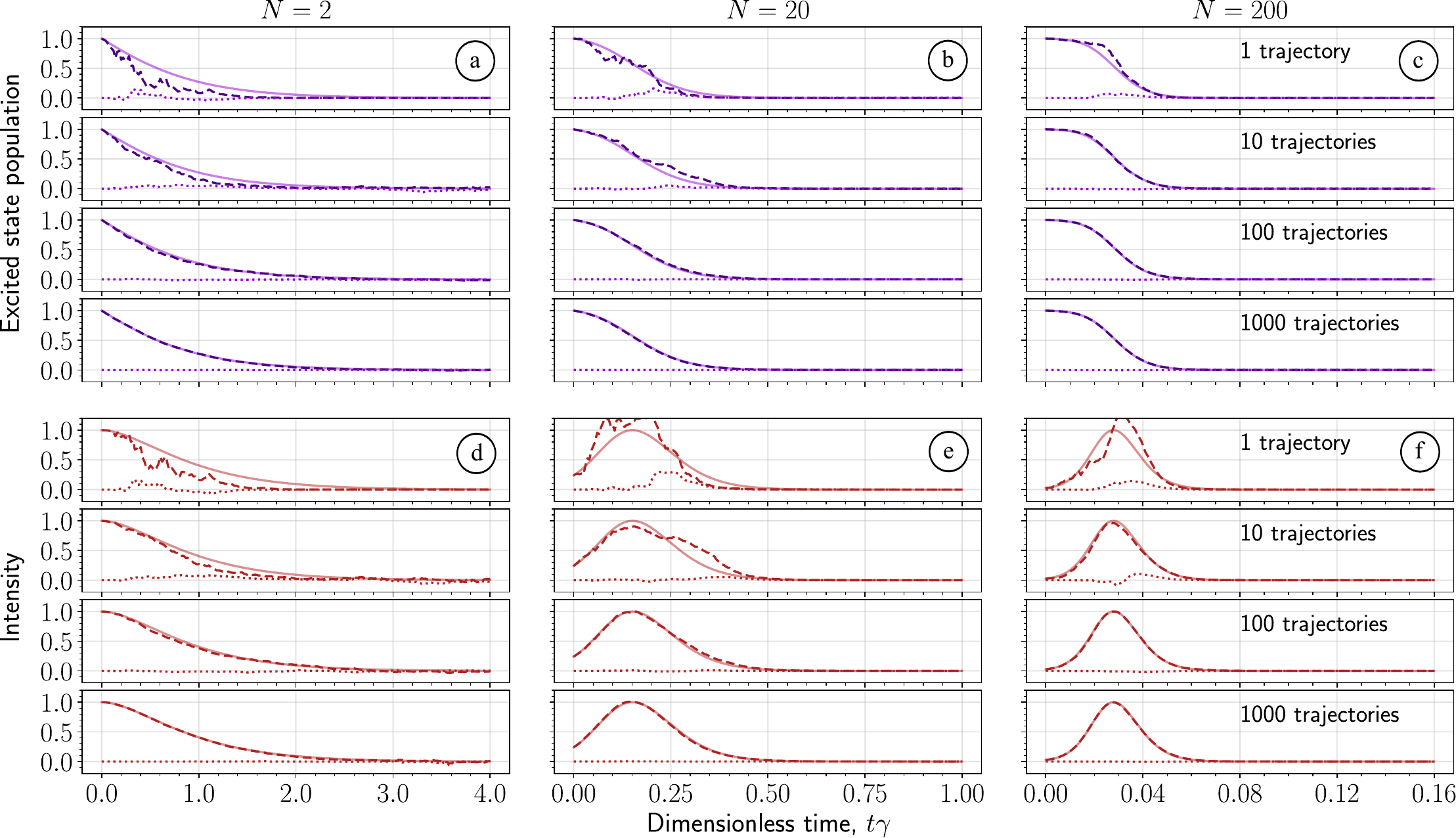}
    \caption{Convergence of the stochastic averages (dashed lines) to the exact quantum expectation values (solid lines) for different numbers of atoms $N$. The atomic ensemble is initially fully excited, i.e. $\rho_{22}(0) = 1.0$, and other matrix elements are zero. The observables chosen here are the probability of finding an excited atom (a, b, c) and the intensity of the emitted field (d, e, f). Intensity is normalized to the maximum. The dotted lines represent the imaginary parts of corresponding quantities. As expected, they disappear with the increasing number of trajectories.\newline~\newline}
    \label{fig:convergence}
\end{figure*}

\section{Numerical analysis\label{sec: Numerical analysis}}

We illustrate the proposed formalism through a series of numerical examples. The deterministic components of the stochastic equations are numerically integrated using the adaptive step-size method \texttt{Tsit5}, which is implemented in the \texttt{DifferentialEquations.jl} library \cite{Rackauckas2017}. The noise components are integrated using the Euler-Maruyama method \cite{milstein_stochastic_2004, Maruyama1955}. {We use $10^5$ stochastic trajectories to construct statistical averages.} {The simulations based on the stochastic formalism are compared with those based on the methodology presented in Ref. \cite{Sukharnikov2023}.} For both methods, the maximum allowed timestep was limited to {$T_\text{max}/ 10^{4}$}. Here, $T_\text{max}$ represents the last point on the dimensionless time grid.

 \begin{figure*}[t!]
    \centering
    \includegraphics[width = \linewidth]{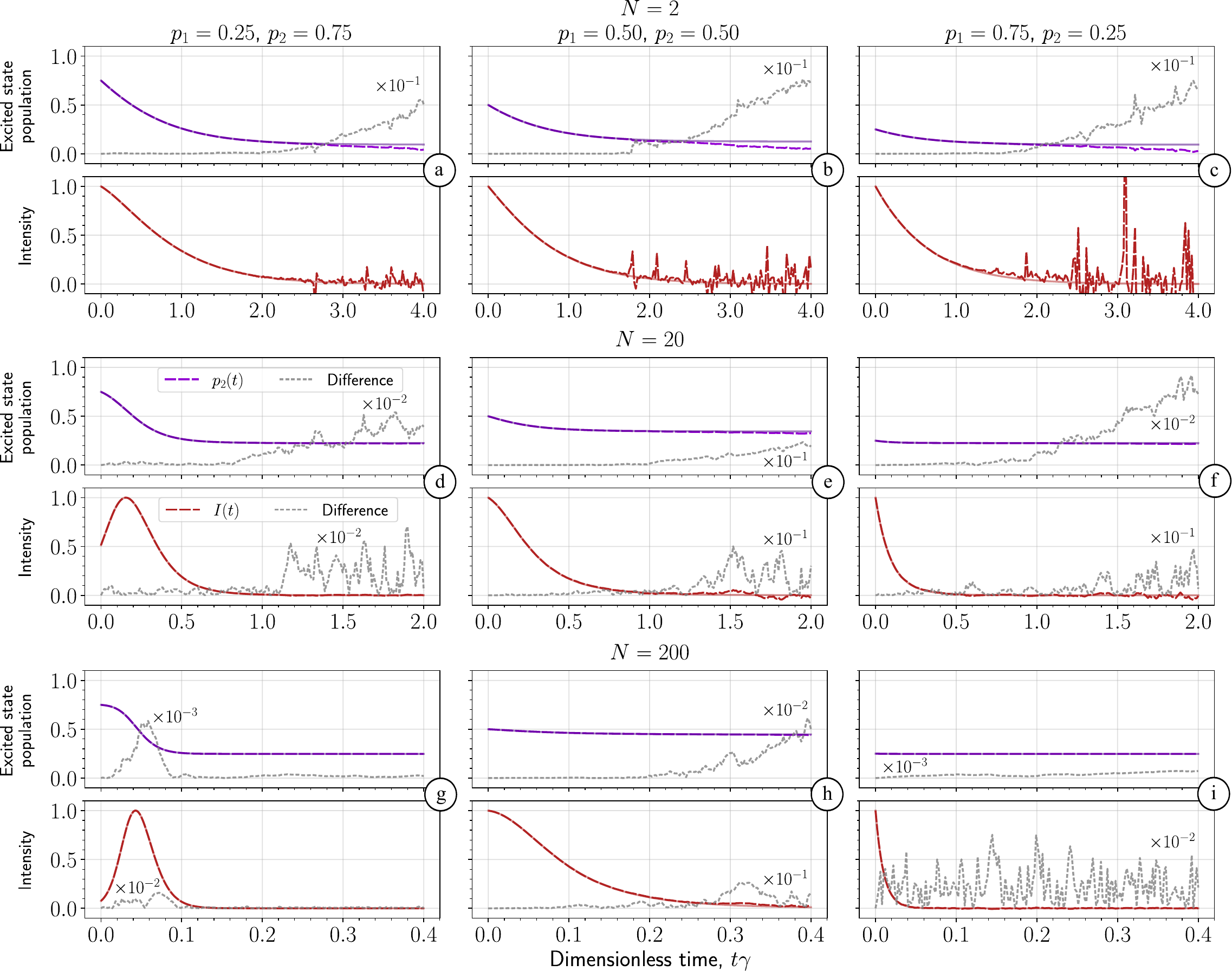}
    \caption{The excited state population and emission intensity plotted for different numbers of atoms $N$ and varying initial conditions. The semi-transparent lines correspond to the quantum expectation values, while the opaque lines represent the stochastic averages. {The absolute differences between these values are indicated by gray dotted lines. We label some plots with $\times 10^{-1}$, $\times 10^{-2}$, $\times 10^{-3}$ to highlight that the corresponding difference values should be multiplied by this factor. }  { We excluded the following numbers of unstable trajectories: (a) 131 (0.13\%); (b) 178 (0.18\%); (c) 170 (0.17\%); (d) 32 (0.03\%); (e) 109 (0.11\%); (f) 40 (0.04\%); (h) 17 (0.02\%).}\newline~\newline}
    \label{fig:various_initial}
\end{figure*}

{ The dynamics of the stochastic density matrix $\rho_{pq}(t)$ is defined by Eq.~\eqref{eq: bloch equations} with noise terms given in Eq.~\eqref{eq: noise decomposition}}. As proposed in  in Sec. \ref{sec: stochastic gauges}, we apply stochastic gauges to mitigate the instabilities. We modify the equations according to Eq. \eqref{eq: field modification}, namely by making the deterministic part of the field Hermitian. This adjustment requires the introduction of an additional variable $C_0(t)$, which re-weights the trajectories, as shown in Eq. \eqref{eq: weighted stochastic density matrix}. Additionally, we employ the diffusion gauge presented in Eq.~(\ref{eq: specific form of the elementary noise terms}) and Appendix~\ref{app:Diffusion gauge} to suppress the increase of non-Hermitian dipole moment components. { As explained at the end of Sec. \ref{sec: stochastic gauges}, a stochastic trajectory is removed from the statistical sample if it exhibits diverging behavior, with its weight coefficient $\Omega(t)$ exceeding $e^5$ or any density matrix element growing above 100. The number of excluded trajectories is given in the caption of each figure.}

Before we proceed with numerical simulations, let us link the expectation values of quantum-mechanical operators with the respective stochastic variables. Specifically, we will examine the average populations of atomic levels using the following expression:
\begin{equation*}
    p_q(t)=\frac{1}{N}\sum_a\text{Tr}(\hat{\sigma}_{a,qq}\hat{\rho}(t))=\langle\Omega(t)\rho_{qq}(t)\rangle,
\end{equation*}
where Eq.~(\ref{eq: weighted stochastic density matrix}) has been utilized.

In compact systems, the field properties can be expressed through the atomic operators and, consequently, through the associated stochastic variables. Up to an insignificant factor, the intensity of the emission polarized along the $\alpha$-axis is given by the product of collective dipole moments:
\begin{subequations}
\label{eq: intensity of the emission}
\begin{equation}
    I_\alpha(t) =\text{Tr}\left[\hat{P}_\alpha^{(-)}\hat{P}_\alpha^{(+)}\hat{\rho}(t)\right].
\end{equation}
Utilizing Eq.~(\ref{eq: weighted stochastic density matrix}), we express the intensities $I_\alpha(t)$ in terms of the stochastic variables $\rho_{pq}(t)$:
\begin{multline}
I_\alpha(t) = N \sum_{\mathclap{e_1,e_2,g}} \; d_{e_1g,\alpha}\, d_{ge_2,\alpha} \, \langle \Omega(t)\rho_{e_2e_1}(t) \rangle \\ +N(N-1) \sum_{\mathclap{\substack{e_1,e_2\\g_1,g_2}}} \, d_{e_1g_1,\alpha}\, d_{g_2e_2,\alpha} \big\langle \Omega(t){\rho}_{g_1e_1}(t) \, {\rho}_{e_2g_2}(t) \big\rangle.
\end{multline}
\end{subequations}
The full intensity is found by summing all the components.

\subsection{Cooperative emission of two-level atoms}
\label{Cooperative emission of two-level atoms}

Let us revisit the example of identical two-level atoms collectively interacting with their own field. The ground state manifold $\{\ket{g}\}$ collapses to a single state $\ket{1}$, and the excited state manifold $\{\ket{e}\}$ reduces to a single state $\ket{2}$.

When the ensemble starts from the fully excited state, characterized by $\rho_{22}(0) = 1$, with all other matrix elements set to zero, the phenomenon of superradiance is observed. Fig. \ref{fig:comparison} shows the excited state population and emission intensity for different numbers of atoms $N$. We exclude the single-atom case $N = 1$ since it does not require the extension of the ansatz \eqref{eq: classical sampling} beyond the single-particle density matrix. Thus, the noise terms are unnecessary in this case. Increasing the value of $N$ leads to faster depopulation of the excited state and a narrower, more pronounced peak in intensity. As depicted in Figs.~\ref{fig:comparison}, the discrepancy between stochastic averages and quantum expectation values remains less than {$1.0\%$}. {Panels (a) and (d) show that the difference is higher for smaller \( N \). We attribute this increase to the non-linear noise terms that become comparable to the deterministic parts for small $N$.} { Moreover, it can lead to unstable trajectories and small spikes in intensities, as observed in the simulations shown in Figs.~\ref{fig:comparison} (a, d). }

In Fig. \ref{fig:comparison} we used $10^5$ stochastic realizations. In practice, much fewer trajectories are required for the convergence of selected observables. Fig. \ref{fig:convergence} demonstrates the convergence of population and intensity for different numbers of stochastic realizations. For qualitative analysis, averaging over $10^2$ trajectories is enough, while averaging over $10^3$ trajectories already gives accurate averages. { Based on our experience, $1000$ trajectories are also sufficient to achieve good accuracy for other systems discussed later.} 
 In addition to the direct comparison with quantum averages, we use another criterion of convergence: imaginary parts of such observables as populations or intensities should vanish after averaging. For a single trajectory, the imaginary part is comparable to the real part, thus it does not have physical meaning. Statistical averages represent observables only after averaging over a significant amount of trajectories. 

There is a special case when the initial state is statistically mixed. As shown in Ref. \cite{Sukharnikov2023}, if the system is prepared in the state without coherences $\rho_{12}(0) = \rho_{21}(0) = 0$ and only with the diagonal elements:
\begin{align}\label{eq: statistically mixed state}
    & \rho_{11}(0) = p_1, 
    & \rho_{22}(0) = p_2,
\end{align}
with $p_2 < 1$, the collective emission process {becomes weaker}. The ensemble reaches a steady state with a nonzero probability of finding an excited atom, namely $\langle \rho^{(ss)}_{22} \rangle> 0$, where $ss$ stands for steady state\footnote{Here, in averages we dropped out the weight function for brevity.}. In this steady state, the field intensity, as defined in Eq.~\eqref{eq: intensity of the emission}, is zero, implying that:
\begin{equation}
\label{eq: non-hermitian 1}
    \langle\rho^{(ss)}_{22} \rangle + (N - 1) \, \langle \rho^{(ss)}_{12} \, \rho^{(ss)}_{21} \rangle = 0.
\end{equation}
This, in turn, implies that $\langle \rho^{(ss)}_{12} \, \rho^{(ss)}_{21} \rangle < 0$, a condition that can only be met when the dipole moments exhibit significant non-Hermitian behavior at the level of individual stochastic realizations. Our gauges aim to minimize the non-Hermitian components, and finding appropriate gauging to account for this particular case remains a separate challenge. And indeed, this special case is not fully reproduced by the stochastic equations, see Fig. \ref{fig:various_initial}. Panels (a-c) show that, for a small number of atoms $N = 2$, the excited state population does not converge to the correct curve, while the intensity exhibits slow convergence and noisy behavior at later time moments. As demonstrated in Fig. \ref{fig:various_initial}~(d-i), the situation improves when the number of atoms increases. { Specifically, the percentage of unstable trajectories becomes lower. Additionally, the absolute difference between populations and intensities, based on stochastic and full quantum-mechanical approaches, decreases. 

Note that the case of \(p_1 = p_2 = 0.5\) consistently exhibits worse performance in terms of absolute differences and numbers of unstable trajectories for any number of atoms.} In Ref. \cite{Sukharnikov2023}, an analytical expression for the steady-state density matrix was found. Initial conditions enter this expression as a single parameter $(p_1 \, p_2) \leq 1/4$. The larger this parameter, the stronger the population trapping effect becomes, {making it more challenging to reproduce the correct curves using the stochastic methodology.}


 \begin{figure*}[t!]
    \centering
    \includegraphics[width = \linewidth]{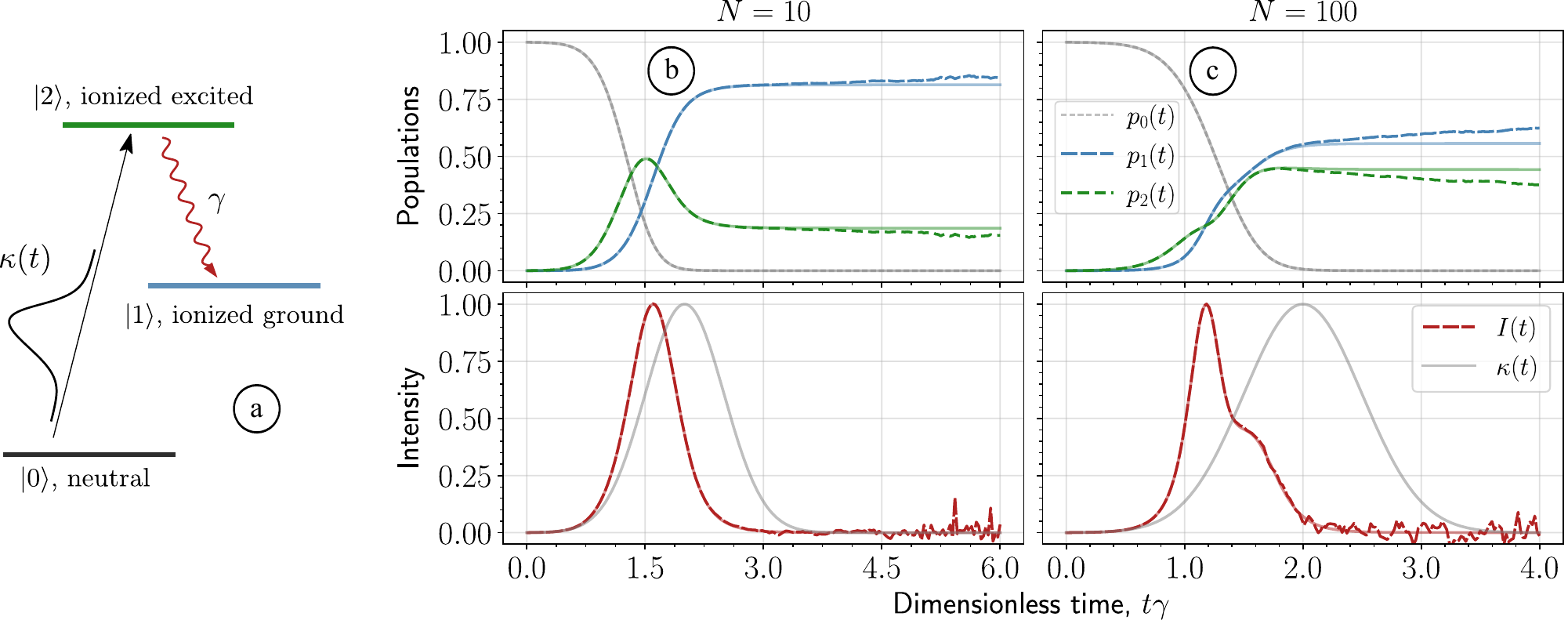}
    \caption{(a) Level structure of the pumped two-level atoms. Neutral atoms are photoionized by a pump pulse with a profile $\kappa(t)$. Excited ions relax via collective emission $2\rightarrow 1$ with a rate $\gamma$. We compare the stochastic (opaque lines) and quantum (semi-transparent lines) expectation values for two cases: $N = 10$ (b) and $N = 100$ (c). We have selected a Gaussian envelope for the pump, $\kappa(t) = I_p \exp\left[{-\frac{(t - t_0)^2}{2\tau^2}}\right] / \sqrt{2\pi \tau^2}$, with the following parameters used for calculations: $I_p = 10$, $t_0 = 2.0 / \gamma$, and $\tau = 0.5 / \gamma$. { We excluded 199 (0.20\%) (b) and 3871 (3.87\%) (c) unstable trajectories.}\newline}
    \label{fig: quasi lambda pumped}
\end{figure*}
 \begin{figure*}[t!]
    \centering
    \includegraphics[width = \linewidth]{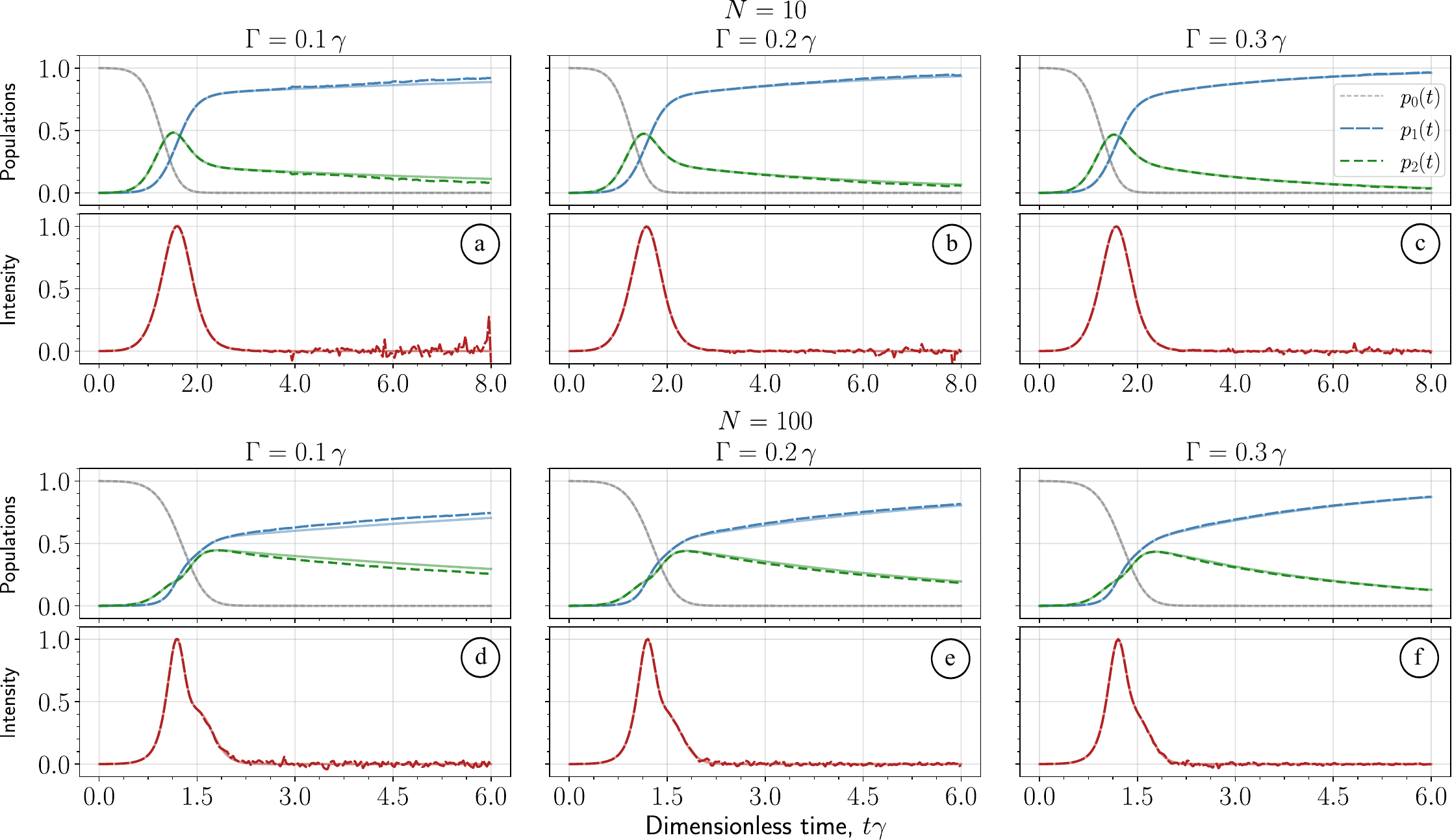}
    \caption{Regularization of the incoherently pumped system from Fig. \ref{fig: quasi lambda pumped} by introducing an additional non-radiative dissipation channel $2\rightarrow 1$ with a rate $\Gamma$. The semi-transparent lines correspond to quantum expectation values, while the opaque lines represent stochastic averages. {  We neglected (a) 268 (0.27\%); (b) 119 (0.12\%); (c) 59 (0.06\%); (d) 2699 (2.70\%); (e) 1061 (1.06\%); and (f) 493 (0.49\%) unstable trajectories.}\newline~~}
    \label{fig: quasi lambda gamma}
\end{figure*}

The population trapping effect occurs due to the assumption that all the atoms experience the same field. This field is immediately updated according to the current value of the dipole moments. At some point, the emission and absorption processes balance each other, and the ensemble does not relax to the ground state but rather evolves into a quasi-stationary state. 

\subsection{Incoherent pumping}\label{Sec: quasi lambda}

In a more realistic situation, the excitation of the ensemble is not instantaneous. The system's initial condition may be prepared by continuous incoherent pumping, as in x-ray lasing experiments \cite{2014'Weninger, Rohringer2012}. A pump pulse ionizes neutral atoms, opening the lasing transition in the ionized atoms, see Fig. \ref{fig: quasi lambda pumped} (a) for a sketch of the level structure. In Ref. \cite{Sukharnikov2023}, it was shown that if the system is pumped incoherently, it reaches a steady state similar to the ones in Fig. \ref{fig:various_initial}, in which the atoms are not fully relaxed. 


At the level of stochastic equations, the additional variable describing the population of the neutral state is required, denoted by $\rho_{00}(t)$. This state is coupled to the excited state only through the incoherent pumping $0 \rightarrow 2$ with the rate $ \kappa(t)$. The time dependence of $\kappa(t)$ defines the pump profile. Pumping results in an additional term \newpage~\newpage~\newpage\noindent in equation for $\rho_{22}(t)$:
\begin{equation*}
    \dot{\rho}_{22}(t) = ... + \kappa(t) \, \rho_{00}(t),
\end{equation*}
while the new variable satisfies the stochastic equation:
\begin{multline}\label{eq: rho00 equation}
    \dot{\rho}_{00}(t) = -\kappa(t) \rho_{00}(t) \\
    - \rho_{00}(t) \sqrt{\dfrac{\gamma}{2}} \sum_{e,g}\left\{ \mathbf{d}_{eg} \rho_{ge}(t) \mathbf{f}^\dagger(t) + \mathbf{d}_{ge} \rho_{eg}(t) \mathbf{g}^\dagger(t) \right\}.
\end{multline}

 In our simulations, we have used a Gaussian pump profile. The stochastic averages are depicted in Fig. \ref{fig: quasi lambda pumped}. {Although the superradiant dynamics is accurately reproduced,} we encounter the same issue as in Fig. \ref{fig:various_initial}, where the steady states are captured incorrectly. { While the populations do not maintain constant values, the intensities show slow convergence and small spikes at the end of the evolution time.} For a larger number of atoms $N = 100$ in panel (c), the situation does not improve, in contrast to Fig. \ref{fig:various_initial} (g, i). {Additionally, the number of unstable trajectories increases.} Perhaps, the continuous pumping process generates a more intricate steady state, posing challenges for accurate reproduction by our stochastic formalism.

As pointed out in Ref. \cite{Sukharnikov2023}, additional dissipation channels disrupt the formation of the steady states. Ref.~\cite{Sukharnikov2023}~assumed the Meitner-Auger decay of the excited state. Alternatively, one could consider non-radiative dissipation to the ground state ($2 \rightarrow 1$) with a rate $\Gamma$ as follows:
\begin{align*}
    & \dot{\rho}_{11}(t) = \ldots + \Gamma \rho_{22}(t),
    & \dot{\rho}_{22}(t) = \ldots - \Gamma \rho_{22}(t), 
    \\
    & \dot{\rho}_{12}(t) = \ldots - \dfrac{\Gamma}{2} \rho_{12}(t),
    & \dot{\rho}_{21}(t) = \ldots - \dfrac{\Gamma}{2} \rho_{21}(t).
\end{align*}
Fig. \ref{fig: quasi lambda gamma} illustrates the impact of non-radiative dissipation on the discrepancies observed in Fig. \ref{fig: quasi lambda pumped}. For $N=10$ {and $100$}, we increase the dissipation rate $\Gamma$ from {$0.1\gamma$ (panel (a, d))} to {$0.3\gamma$ (panel (c, f)), which gradually} improves the performance of the stochastic method. {Populations show better agreement with the full quantum simulations, while intensities exhibit better convergence and fewer spikes. Additionally, the number of unstable trajectories decreases.} Based on our experience, systems with a larger number of atoms may require a higher dissipation rate. In the given examples, the additional dissipation with the rate of the same order of magnitude as spontaneous emission, i.e., $\Gamma \sim \gamma$, is sufficient for regularization.

 \begin{figure*}[t!]
    \centering
    \includegraphics[width = \linewidth]{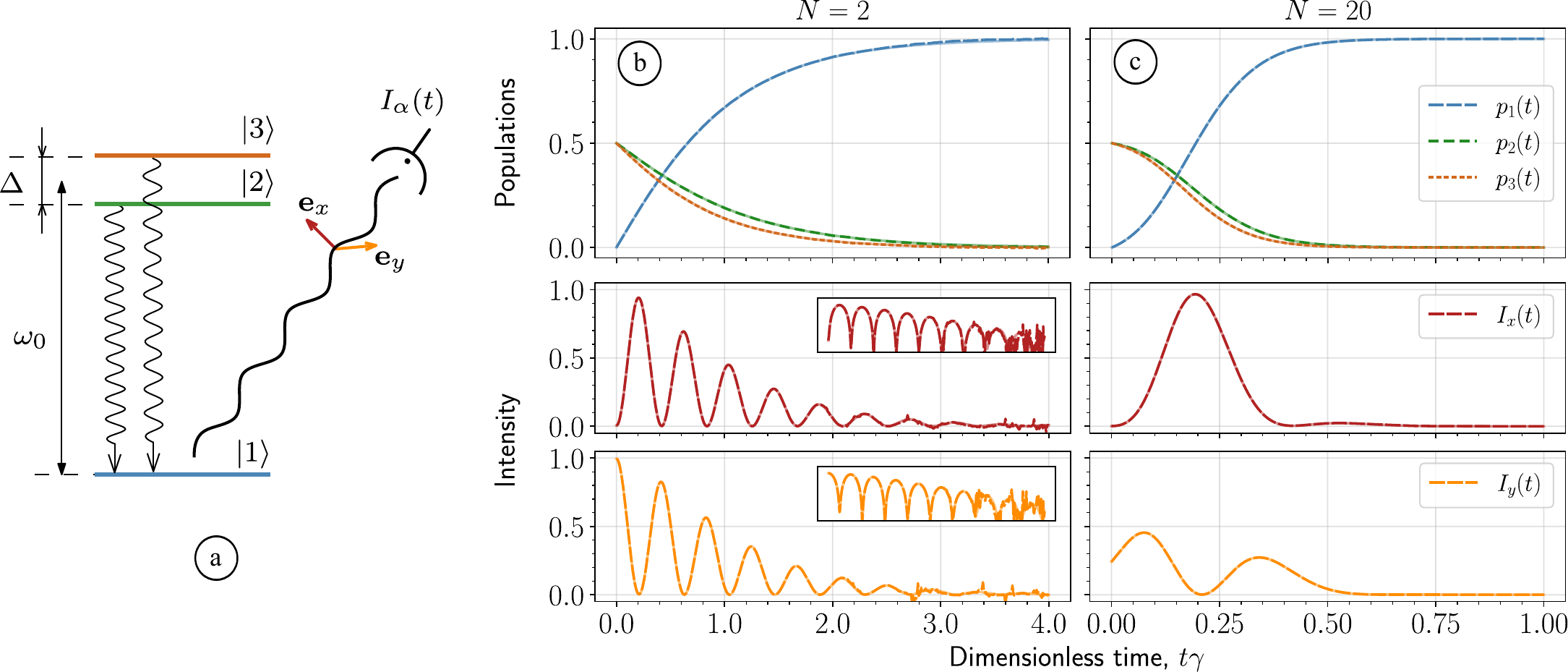}
    \caption{Quantum beats in a $V$-type system, depicted on the left (a), calculated for $N = 2$ (b) and $N = 20$ (c) atoms. In both cases, we have taken $\Delta = 15 \, \gamma$, and plotted populations (upper row) and normalized intensities {(lower rows)} of the field for both polarizations. For intensity curves (b), we give the same plots in a logarithmic scale in small boxes. The intensities of different polarization components are normalized to the maximum full intensity. The semi-transparent lines correspond to quantum expectation values, while the opaque lines represent stochastic averages. We { omitted $47$ (0.05\%)} diverging trajectories for the case of $N = 2$ (b).}
    \label{fig: V beats}
\end{figure*}

 \begin{figure*}[t!]
    \centering
    \includegraphics[width = \linewidth]{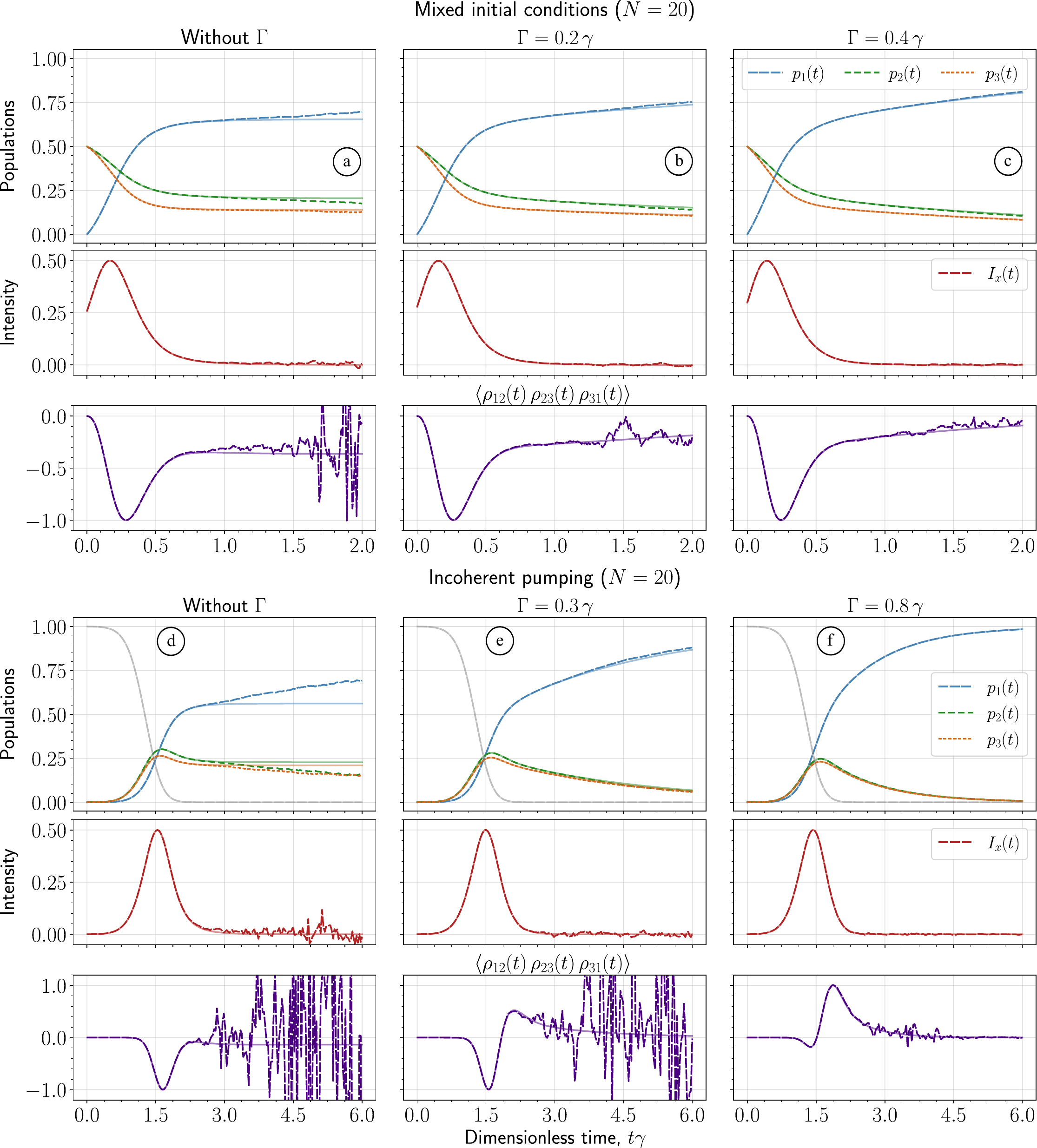}
    \caption{The dynamics of the ensemble of $N = 20$ atoms with the $V$-type level structure depicted in Fig. \ref{fig: V beats} (a). In order to mitigate the discrepancies, we introduce additional non-radiative dissipation channels from the excited states to the ground state ($2,3 \to 1$) with a rate $\Gamma$ (see Eqs.~(\ref{eq: additional decay V})). The semi-transparent lines correspond to quantum expectation values, while the opaque lines represent stochastic averages. In the upper row (a, b, c), atoms start from the mixed state without coherence $\rho_{22}(0) = \rho_{33}(0) = 0.5$, the rest is zero. In the lower row (d, e, f), atoms are incoherently pumped, and similar issues with steady states arise. We chose the same profile as in Fig. \ref{fig: quasi lambda pumped} for the pump. The gray line demonstrates the evolution of $\rho_{00}(t)$. Overall we omitted { (a) 427 (0.43\%); (b) 174 (0.17\%); (c) 63 (0.06\%); (d) 2503 (2.50\%); (e) 239 (0.24\%) and (f) 4 (0.004\%) unstable trajectories}. { Both polarization  components of intensity exhibit identical profiles, and we depict only one of them.}}
    \label{fig: V gammas}
\end{figure*}

 \begin{figure*}[t!]
    \centering
    \includegraphics[width = \linewidth]{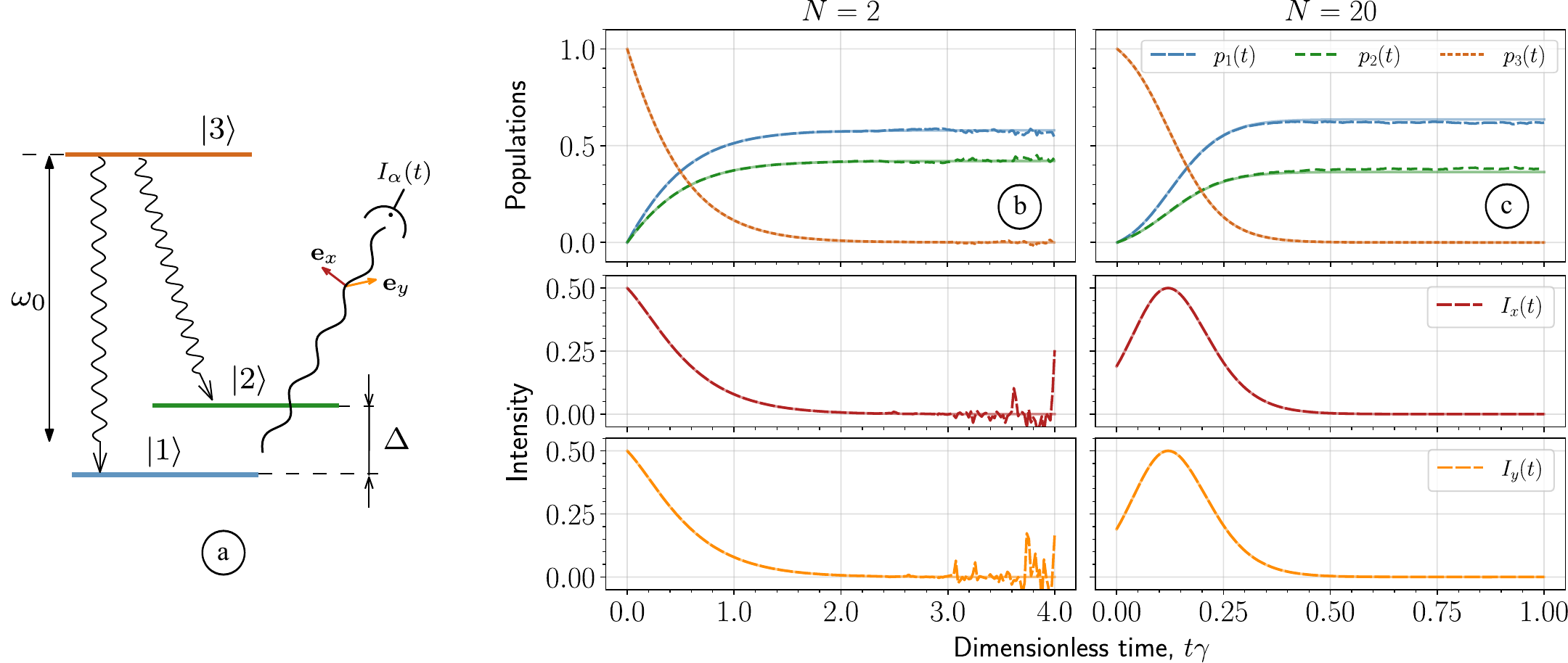}
    \caption{Lasing from an ensemble of atoms with a $\Lambda$-type level structure, as depicted in (a). The separation between ground states $\Delta$ is much smaller than the center frequency $\omega_0$. We present the evolution of populations and intensity components for $N = 2$ (b) and $N = 20$ (c) atoms. Intensity components are normalized to the maximum full intensity. The semi-transparent lines correspond to quantum expectation values, while the opaque lines represent stochastic averages. {We neglected $383$ (0.39\%) (b) and $2229$ (2.23\%) (c) unstable realizations}.}
    \label{fig: lambda}
\end{figure*}

In conclusion, the ratio between timescales of superradiance, pumping, and dissipation directly influences the formation of steady states. Notably, when these steady states are less prominent, the stochastic formalism consistently yields accurate averages.

\subsection{Quantum beats in $V$-system}\label{Quantum beats in V system}

So far, we have considered only models with lasing between two levels. To demonstrate that our formalism correctly captures many-level effects, we consider a $V$-type configuration with two excited states $\ket{2}, \ket{3}$ and a single ground state $\ket{1}$, as shown in Fig. \ref{fig: V beats} (a). The energy gap between excited states $\Delta$ is much smaller than the center frequency $\Delta \ll \omega_0$. Fluorescence from emitters with such a level structure may exhibit quantum beating, a fundamental quantum phenomenon that has been observed in various spectral ranges, including optical \cite{1973'Aleksandrov, 1990'Bitto}, XUV \cite{2020'Hikosaka, LaForge2020}, and x-rays \cite{1986'Gerdau}. In the context of collective emission, quantum beating is superimposed with superfluorescent behavior \cite{1977'Vrehen, 1994'Bartholdtsen}. Here, we consider superfluorescence in Helium gas under presence of a weak magnetic field, as in Ref. \cite{LaForge2020}. We consider transitions from states $2$ to $1$ and $3$ to $1$ with slightly different strengths:
\begin{align*}
& \mathbf{d}_{31} = \dfrac{d_{31}}{\sqrt 2} \left( \mathbf{e}_x - i \mathbf{e}_y \right),
& \mathbf{d}_{21} = \dfrac{d_{21}}{\sqrt 2} \left( \mathbf{e}_x + i \mathbf{e}_y \right),
\end{align*}
where $|d_{31}|^2 = 1$ and $|d_{21}|^2 = 0.75$. The transition between the excited states is forbidden, namely $\mathbf{d}_{32} = 0$. If we assume that each atom in the ensemble is initially prepared in a coherent superposition of excited states, such as:
\begin{equation*}
    |\psi\rangle_a = \dfrac{|2\rangle_a - | 3 \rangle_a}{\sqrt{2}},
\end{equation*}
we can observe quantum beats in the intensity of both the $x$- and $y$-components of the field \cite{Haroche1976}. In this scenario, the initial collective state of the ensemble is separable, following Eq. \eqref{eq: classical sampling}, and each atom is characterized by a single-particle density matrix with the following components:
\begin{align*}
& \rho_{22}(0) = \rho_{33}(0) = 0.5,
& \rho_{23}(0) = \rho_{32}(0) = -0.5,
\end{align*}
while all other matrix elements are zero.

As demonstrated in Fig. \ref{fig: V beats} (b, c), our formalism reproduces the quantum beats. For $N = 2$ atoms {in panel (b)}, we additionally plot intensity curves in the logarithmic scale (a small box at the top right corner). {The stochastic formalism agrees well with the full quantum-mechanical calculations. Only the low-amplitude oscillations are not reproduced, because they are at the level of statistical fluctuations.}

Let us also analyze a system starting from a statistical mixture. We focus on the most challenging scenario where the initial state has no coherences, and the excited states are statistically equally populated:
\begin{align*}
    & \rho_{22}(0) = \rho_{33}(0) = 0.5,
    & \rho_{23}(0) = \rho_{32}(0) = 0.0.
\end{align*}
In this case, there are no quantum beats in the intensity, and the ensemble evolves into a nontrivial steady state. { 
As depicted in Fig. \ref{fig: V gammas} (a), our formalism does not entirely reproduce this steady state for longer evolution times.} 

With multi-level atoms, we can construct another class of observable, namely a three-operator correlator:
\begin{equation*}
\begin{split}
    \langle \rho_{12}(t) \, \rho_{23}(t) \, \rho_{31}(t)\rangle \\ = \dfrac{(N-3)!}{N!} \sum_{\mathclap{\mu_1 \not = \mu_2 \not = \mu_3}} \mathrm{Tr} \left( \hat{\sigma}_{\mu_1, 13} \, \hat{\sigma}_{\mu_2, 32} \, \hat{\sigma}_{\mu_3, 21} \, \hat{\rho}(t) \right).
\end{split}
\end{equation*}
Such correlators are often factorized in semi-classical and approximated approaches \cite{2019'Benediktovitch}. However, when there is no initial coherence, any factorization of this operator results in zero. Hence, it is important to demonstrate how our formalism reproduces such correlators. As shown in Fig. \ref{fig: V gammas} (a), the convergence of the three-operator correlator becomes problematic {only when} the system approaches the steady state.

 \begin{figure*}[t!]
    \centering
    \includegraphics[width = \linewidth]{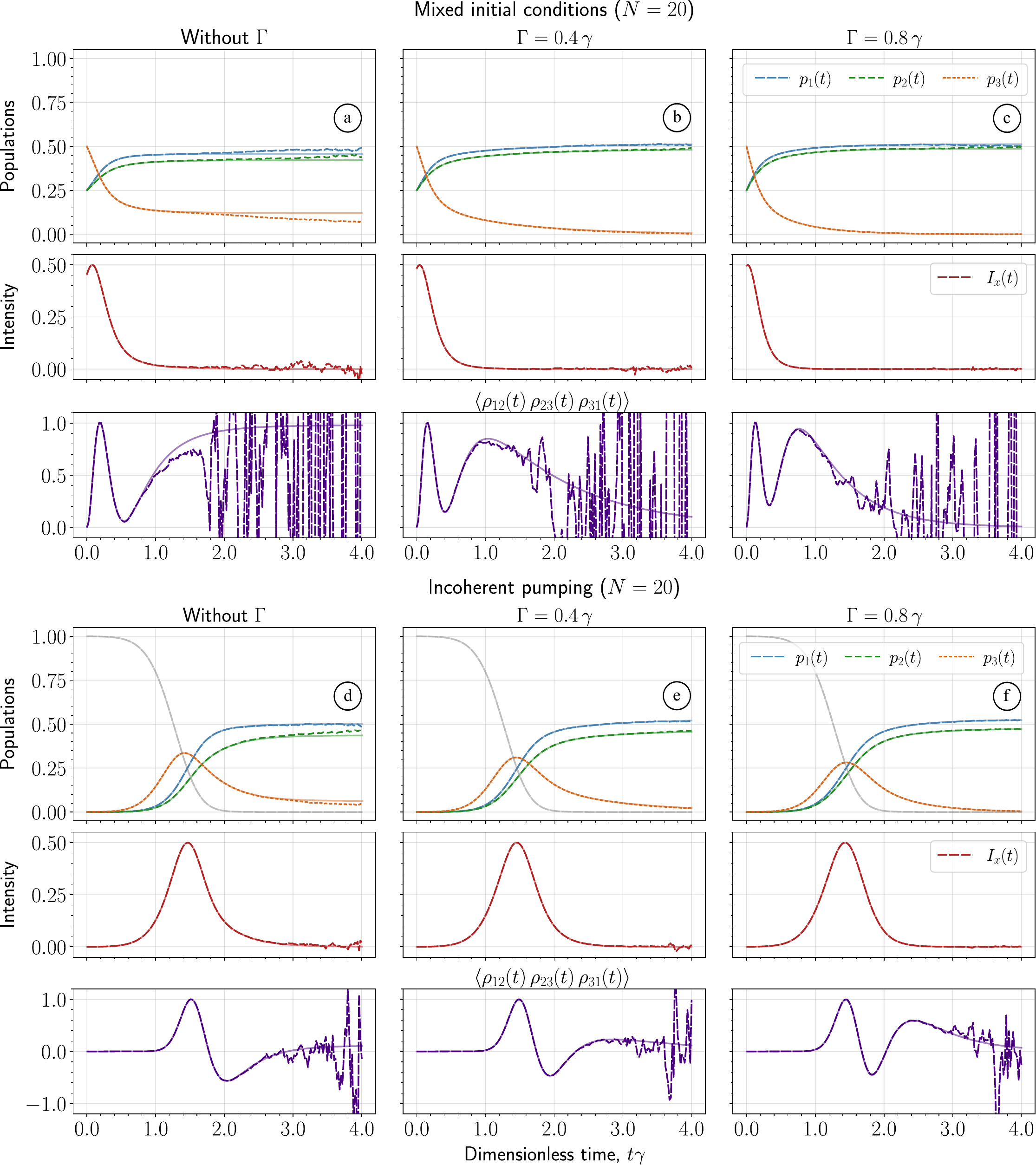}
    \caption{The dynamics of an ensemble of \(N = 20\) atoms with the \(\Lambda\)-type level structure depicted in Fig. \ref{fig: lambda} (a). The figure composition is the same as in Fig. \ref{fig: V gammas}. Semi-transparent lines represent quantum expectation values, while opaque lines show stochastic averages. The upper row (a, b, c) considers atoms starting from the mixed initial state \(\rho_{11}(0) = \rho_{22}(0) = 0.25\) and \(\rho_{33}(0) = 0.50\), with the rest being zero. The lower row (d, e, f) shows atoms being incoherently pumped. To address convergence issues, non-radiative damping of the excited state at rate \(\Gamma\) is introduced (see Eqs.~(\ref{eq: additional decay Lambda})). The gray line shows the evolution of \(\rho_{00}(t)\). {Unstable realizations omitted: (a) 2383 (2.38\%); (b) 985 (0.99\%); (c) 735 (0.74\%); (d) 1550 (1.55\%); (e) 351 (0.35\%); (f) 159 (0.16\%). Both polarization components of intensity exhibit identical profiles, so only one is depicted.}}
    \label{fig: lambda gammas}
\end{figure*}

In reality, the atoms are not pumped instantaneously. To simulate the effect of a pump pulse, we introduce an additional level $|0\rangle$ described by $\rho_{00}(t)$, as in Sec. \ref{Sec: quasi lambda}. All atoms start from this state and are incoherently pumped to the excited states according to:
\begin{align*}
    &\dot{\rho}_{ee}(t) = ... + \dfrac{\kappa(t)}{2} \, \rho_{00}(t),
    & e = 2,3,
\end{align*}
while $\rho_{00}(t)$ satisfies Eq.~\eqref{eq: rho00 equation}. The simulations in Fig.~\ref{fig: V gammas}~(d) reveal that the convergence problems are more pronounced when the system is pumped. {Specifically, the intensity curves have spikes and the three-operator correlator does not converge after the emission peak around $t\gamma \approx 2.0$. This is due to the stronger influence of steady states since more population is trapped in the excited states compared to the case without pumping.}

We regularize these issues by introducing additional non-radiative damping of excited states to the ground states with a rate $\Gamma$:
\begin{subequations}
\label{eq: additional decay V}
\begin{align}
    &\dot{\rho}_{11}(t) = \ldots + \Gamma \sum_e \rho_{ee}(t)\\
    &\dot{\rho}_{ee}(t) = \ldots - \Gamma \rho_{ee}(t), 
\end{align}
where $e = 2,3$. The coherences decay according to:
\begin{align}
    \dot{\rho}_{1e}(t) =& \ldots - \dfrac{\Gamma}{2} \rho_{1e}(t),\\
    \dot{\rho}_{e1}(t)  =& \ldots - \dfrac{\Gamma}{2} \rho_{e1}(t),\\
    \dot{\rho}_{e_1 e_2}(t)  =& \ldots -\Gamma \rho_{e_1 e_2}(t),
\end{align}
\end{subequations}

In Fig. \ref{fig: V gammas} (b, c) and (e, f), we have found a minimal value of $\Gamma$ required to mitigate the discrepancies. As in the previous section, values of $\Gamma \sim \gamma$ are sufficient. {To stabilize higher-order correlators, larger dissipation rates  are necessary.} With this additional damping, all observables are reproduced for a chosen time range, including the three-operator correlator. { As in simulations illustrated in Fig. \ref{fig: quasi lambda gamma}, larger $\Gamma$ decreases the number of unstable trajectories}. This demonstrates that our formalism goes beyond semi-classical models and, when properly regularized, fully captures the quantum effects of many-body correlations.

\newpage~\newpage~\newpage\noindent 
\subsection{Lasing in $\Lambda$-system}\label{Lasing in Lambda system}

Quantum beats in \(V\)-systems are predicted by semi-classical models and stochastic electrodynamics approaches \cite{scully1999quantum}. However, some semi-classical and stochastic models incorrectly predict quantum beats in \(\Lambda\)-systems \cite{GeaBanacloche1988, scully1999quantum}, with one excited state \(\ket{3}\) and two ground states \(\ket{1}\) and \(\ket{2}\), as shown in Fig. \ref{fig: lambda} (a). To demonstrate our formalism's predictive power as a true quantum model, we study such a \(\Lambda\)-system with orthogonal transition polarizations:
\begin{align*}
& \mathbf{d}_{31} = \dfrac{d_{31}}{\sqrt 2} \left( \mathbf{e}_x - i \mathbf{e}_y \right),
& \mathbf{d}_{32} = \dfrac{d_{32}}{\sqrt 2} \left( \mathbf{e}_x + i \mathbf{e}_y \right).
\end{align*}
Here, $|d_{31}|^2 = 1$ and $|d_{32}|^2 = 0.75$. The transition between the ground states is forbidden.

Our equations predict the absence of intensity beats, which aligns with the results from quantum simulations. In Fig. \ref{fig: lambda} (b, c), we present population and intensity curves for $N = 2$ and $N = 20$. Both polarization components of the field do not show any signs of beating and share the same profile. For this reason, we depict only one polarization component for the further examples. When the system has reached the ground states, where no dynamics is expected, the population curves exhibit a slight deviation from the quantum expectation values {for both $N=2$ and $20$. Additionally, spikes are observed in the intensity curves for \(N=2\). For larger \(N\), the number of unstable trajectories increases. The observed discrepancy suggests that, although the excited state is depopulated, a nontrivial steady state forms, possessing specific correlations between the ground states. }

As in the previous section, we also consider the ensemble prepared in a mixed state without coherences. We assume that all levels are initially populated as follows:
\begin{align*}
    &\rho_{33}(0) = 0.5, 
    &\rho_{22}(0) = \rho_{11}(0) = 0.25,
\end{align*}
and other matrix elements are zero. {The solution reveals that the ensemble does not relax completely to the ground states but evolves into a steady state with some population remaining in the excited state, as in Fig. \ref{fig: lambda gammas} (a). The formation of this state is accompanied by unstable behavior of the three-body correlator, which does not converge after $t\gamma \approx 1.0$.}

The same issues appear when the excited state is incoherently pumped with the following additional term in the equations:
\begin{equation*}
\dot{\rho}_{33}(t) = \ldots + \kappa(t) \, \rho_{00}(t),
\end{equation*}
where $\kappa(t)$ defines the pump profile, and $\rho_{00}(t)$ satisfies Eq. \eqref{eq: rho00 equation}. The final state is also a steady state with a nonzero probability of finding an excited atom, as shown in Fig. \ref{fig: lambda gammas} (d). The stochastic averages converge to incorrect values, and the three-body operator becomes unstable for later time moments.

We attempt to regularize these issues by introducing non-radiative damping of the excited state to the ground states ($3\rightarrow 1, 2$) with a rate $\Gamma$
\begin{subequations}
\label{eq: additional decay Lambda}
\begin{align}
    &\dot{\rho}_{33}(t) =  \ldots - 2\Gamma \rho_{33}(t), \\
    &\dot{\rho}_{gg}(t) =  \ldots + \Gamma \rho_{33}(t), \\
    &\dot{\rho}_{g3}(t) = \ldots - \Gamma \rho_{g3}(t),\\
    &\dot{\rho}_{3g}(t)  = \ldots - \Gamma \rho_{3g}(t),
\end{align}
\end{subequations}
\newpage~\newpage~\newpage \noindent where $g = 1,2$. Consequently, the excited state is depopulated at a rate of $2\Gamma$. {We found the damping rates sufficient to regularize populations and intensities in Fig. \ref{fig: lambda gammas} (b, c) and (e, f). However, the three-operator correlator still does not converge, { although it shows a right trend. A possible} explanation is that the ensemble does not simply evolve into a mixture of ground states. Calculations based on Ref. \cite{Sukharnikov2023} reveal that the system in {Fig.~\ref{fig: lambda gammas} (c)} evolves into a steady state with the following non-zero correlation:
\begin{equation}
\label{eq: non-hermitian 2}
    \langle \rho_{12}^{(ss)} \, \rho_{21}^{(ss)} \rangle < 0,
\end{equation}
where $ss$ stands for steady state. As the product $\rho_{12}^{(ss)} \rho_{21}^{(ss)}$ becomes negative upon averaging, it indicates that the coherences $\rho_{12}(t)$ and $\rho_{21}(t)$ are non-Hermitian at the level of single realizations. {This nontrivial dynamics may cause slow convergence of the three-operator correlators observed in Fig. \ref{fig: lambda gammas}.}  The coherences between the ground states $\rho_{g_1 \not = g_2}(t)$ are not damped by any additional non-radiative decay, which does not prevent slow convergence of the observables involving $\rho_{g_1 \not = g_2}(t)$, such as the analyzed three-particle correlator. Effective regularization would require either stronger decoherence to prevent the buildup of the correlations or the introduction of another dissipation channel for $\rho_{g_1 \not = g_2}(t)$.}

\nocite{*}

\section{Discussion}\label{discussion}


\begin{table*}[t!]
    \centering
    \includegraphics[scale=1]{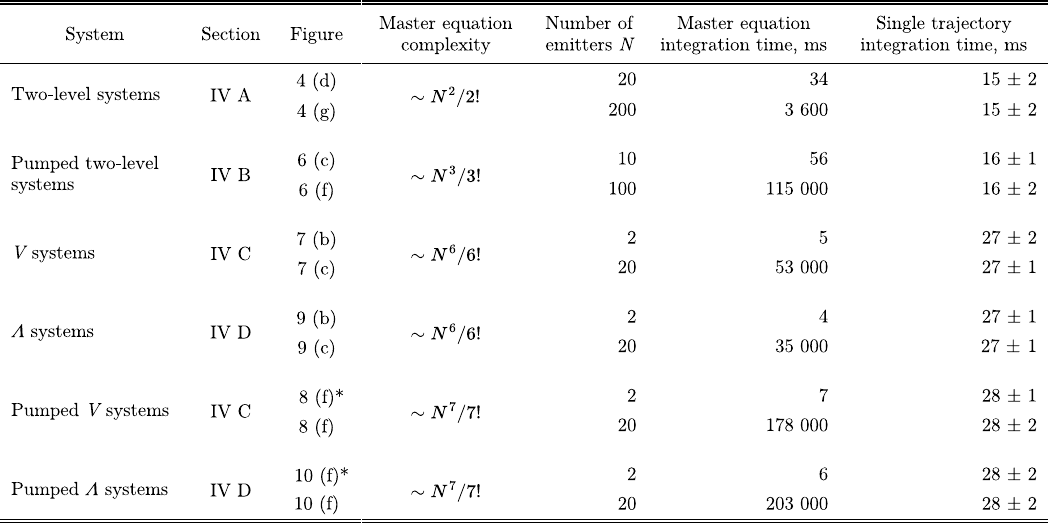}
    \caption{Comparison of computational efforts between the stochastic and quantum-mechanical methods. Each row corresponds to an individual atomic system investigated in the article. The computational time for solving the master equation is averaged over 10 runs. The time required to integrate a single stochastic trajectory (along with one standard deviation) is evaluated based on over 100 realizations. The asterisk (*) in front of the figure number indicates that the simulations are conducted under identical conditions but with a different number of atoms. }
    \label{tab: computation time}
\end{table*}
\subsection{Computational effort}

Firstly, we address the issue of computational efficiency. Table \ref{tab: computation time} provides clear evidence of the effectiveness of the stochastic methodology. For each system analyzed in this article, we compare the average computational time required to simulate a single stochastic trajectory with that of a full quantum-mechanical simulation. The stochastic methodology allows parallelization and remains independent of $N$ across all the examples provided. In contrast,  the polynomial complexity of full quantum-mechanical calculations sharply raises the computational time with an increase in $N$. The performance of the stochastic method is primarily determined by the number of atomic levels $M$, which defines the number of underlying stochastic differential equations $M^2$. As depicted in Fig. \ref{fig:convergence}, convergence of stochastic averages for chosen observables is typically achieved with $10^2$--$10^3$ trajectories. { Based on numerical simulations, pumped $V$ and $\Lambda$ systems also require $10^3$ trajectories to achieve good agreement with the quantum-mechanical simulations}.  {Even without parallelization, the stochastic simulations for the twenty pumped $V$-systems demonstrated in Fig. \ref{fig: V gammas} (c) take between $3$ and $30$ seconds, noticeably less than the $178$ seconds required by full quantum-mechanical calculations.} For larger $N$ and with the implementation of parallelization, the difference becomes even more pronounced.

\subsubsection*{High-order correlation functions}

Another crucial aspect is the convergence of various expectation values. The higher the order of the correlation function, the more pronounced the challenges with convergence become. This issue is consistently observed when the system reaches a special steady state, leading to the effective density matrix $\rho_{pq}(t)$ becoming non-Hermitian at the level of individual realizations. This problem was demonstrated in the context of superfluorescence from two-level atoms in Sec. \ref{Cooperative emission of two-level atoms} and discussed following Eq.~(\ref{eq: non-hermitian 1}). A similar concern was addressed in the context of a $\Lambda$-type system after Eq.~(\ref{eq: non-hermitian 2}) in Sec. \ref{Lasing in Lambda system}. Our drift gauge is designed to ensure the Hermitian behavior of the effective density matrix, and finding appropriate gauging methods to address these steady states remains an open challenge. In our approach, the convergence issue can only be avoided by introducing additional dissipation for the problematic steady states.









\begin{figure*}[t!]
    \centering
    \includegraphics[width = \linewidth]{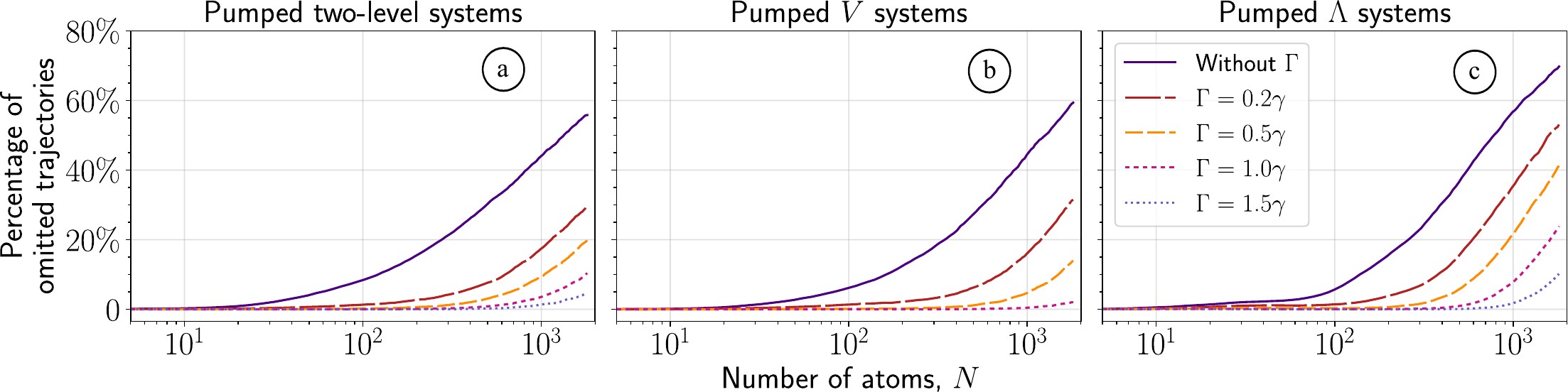}
    \caption{Percentage of omitted trajectories for different level schemes, numbers of emitters and decay rates \(\Gamma\). Panel (a) depicts pumped two-level systems (Sec. \ref{Sec: quasi lambda}), panel (b) pumped \(V\) systems (Sec. \ref{Quantum beats in V system}), and  panel (c) pumped \(\Lambda\) systems (Sec. \ref{Lasing in Lambda system}).}
    \label{fig:gauge_divergences}
\end{figure*}

{ 
\subsection{Unstable trajectories}

Despite the expectation that stochastic gauges can reduce instabilities, almost every numerical example in Sec.~\ref{sec: Numerical analysis} features unstable trajectories, which we associate with the presence of specific steady states. Keeping these unstable trajectories causes large spikes in the temporal profiles of the averages. A larger statistical sample does not smooth the curves; instead, it leads to more diverging trajectories and a higher number of spikes after averaging. Therefore, we omit the unstable trajectories, as detailed in Sec. \ref{sec: stochastic gauges}.

Fig. \ref{fig:gauge_divergences} shows the percentage of omitted trajectories for different level schemes and numbers of emitters, focusing on systems with incoherent pumping. Since incoherent decay processes can mitigate steady-state problems and divergences, each panel in Fig. \ref{fig:gauge_divergences} displays multiple simulations with different decay rates \(\Gamma\).

Qualitatively, each system exhibits similar behavior. Beyond a certain number of atoms, the percentage of omitted trajectories grows faster. This behavior systematically shifts to larger numbers of atoms with an increase in decay rate \(\Gamma\). We expect that the  simulations yield more trustworthy results when the percentage of disregarded trajectories decreases.}

\begin{figure*}[t!]
    \centering
    \includegraphics[width = \linewidth]{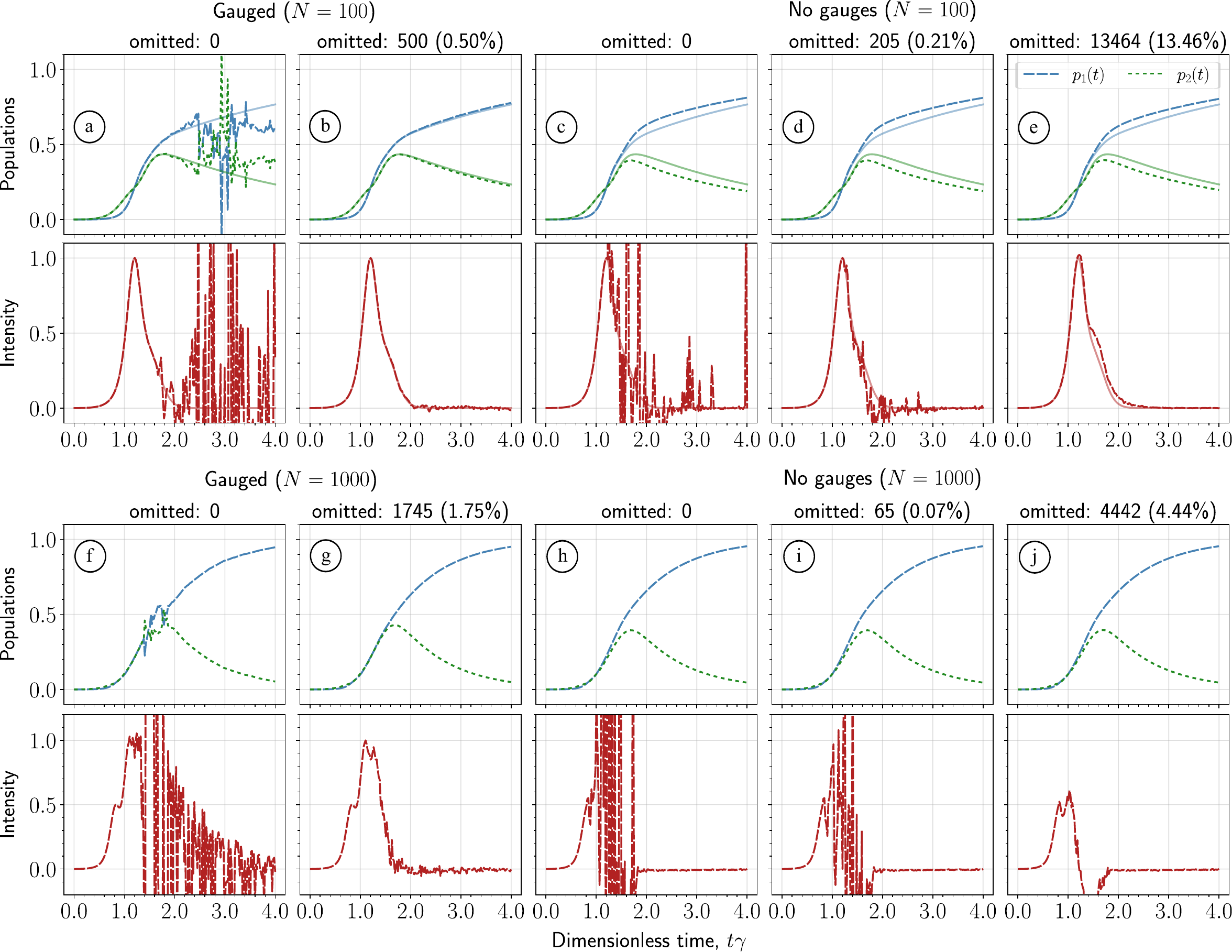}
    \caption{Simulations of population dynamics and intensity profiles for pumped two-level systems (Sec.~\ref{Sec: quasi lambda}), comparing gauged and ungauged equations for \(N = 100\) and \(N = 1000\). The semi-transparent lines correspond to quantum expectation values. The opaque lines represent stochastic averages. The top row shows simulations for \(N = 100\) based on gauged equations with all realizations (a) and with 500 unstable realizations omitted (b), and ungauged equations with all realizations (c), 205 most diverging realizations omitted (d), and all 13 464 diverging realizations omitted (e). The bottom row shows simulations for \(N = 1000\) based on gauged equations with all realizations (f) and with 1745 unstable realizations omitted (g), and ungauged equations with all realizations (h), 65 most diverging realizations omitted (i), and all 4442 diverging realizations omitted (j). }
    \label{fig:gauge_on_and_off}
\end{figure*}

{

\subsection{Stochastic gauging}
\label{sec: stochastic gauges disc}

Since both the gauged and ungauged equations can yield diverging trajectories, comparing simulations based on them could provide valuable insights. We revisit the example of pumped two-level systems discussed in Sec.~\ref{Sec: quasi lambda}. Specifically, we focus on the scenario depicted in Fig. \ref{fig: quasi lambda gamma} (f) with $\Gamma = 0.3\gamma$. Fig.~\ref{fig:gauge_on_and_off} presents two examples, one with \(N=100\) and another with \(N=1000\). For \(N=100\), a quantum-mechanical solution is also included for comparison. Panels (a), (b), (f), and (g) demonstrate simulations based on the gauged equations. The statistical averages shown in panels (a) and (f) are based on all realizations, including the unstable ones, which have large weight coefficients and significantly impact the resulting curves. After excluding $500$ unstable realizations, the curves become much smoother, as shown in panels (b) and (g). In particular, panel (b) shows that removing these unstable realizations leads to a good agreement with the full quantum-mechanical simulations.

Panels (c)-(e) and (h)-(j) showcase simulations based on the ungauged stochastic differential equations. Comparing panels (a) and (c), we observe that convergence issues emerge earlier with the ungauged equations. Panels (d) and (e) illustrate how the averages change as unstable trajectories are gradually removed. Notably, the ungauged equations produce significantly more unstable trajectories than the gauged ones. Furthermore, removing these unstable trajectories does not improve the averages, as evidenced by comparisons with the full quantum-mechanical simulations.

Although we cannot compare the stochastic methodology with full quantum-mechanical simulations for \(N=1000\), a visual inspection suggests that the gauged simulations yield more physically accurate results. Comparing panels (f) and (h), we notice that convergence issues arise earlier with the ungauged equations. Panels (i) and (j) demonstrate that removing unstable trajectories leads to negative intensities when stochastic gauging is not applied. This unphysical behavior does not occur in the gauged simulations shown in panel (g).

The discrepancy between simulations based on the ungauged equations and those based on the full quantum-mechanical approach is not always as drastic as in the example of pumped two-level systems. For instance, superfluorescence in $N=20$ \(\Lambda\)-systems, studied in Sec. \ref{Lasing in Lambda system}, can be accurately modeled using both the gauged and ungauged equations. However, the gauged equations exhibit slightly more unstable behavior due to the influence of the weight coefficient. This suggests that the gauging condition proposed in Sec. \ref{sec: stochastic gauges} can be further refined, which is a topic for future investigations.}

 \begin{figure*}[t!]
    \centering
    \includegraphics[width = \linewidth]{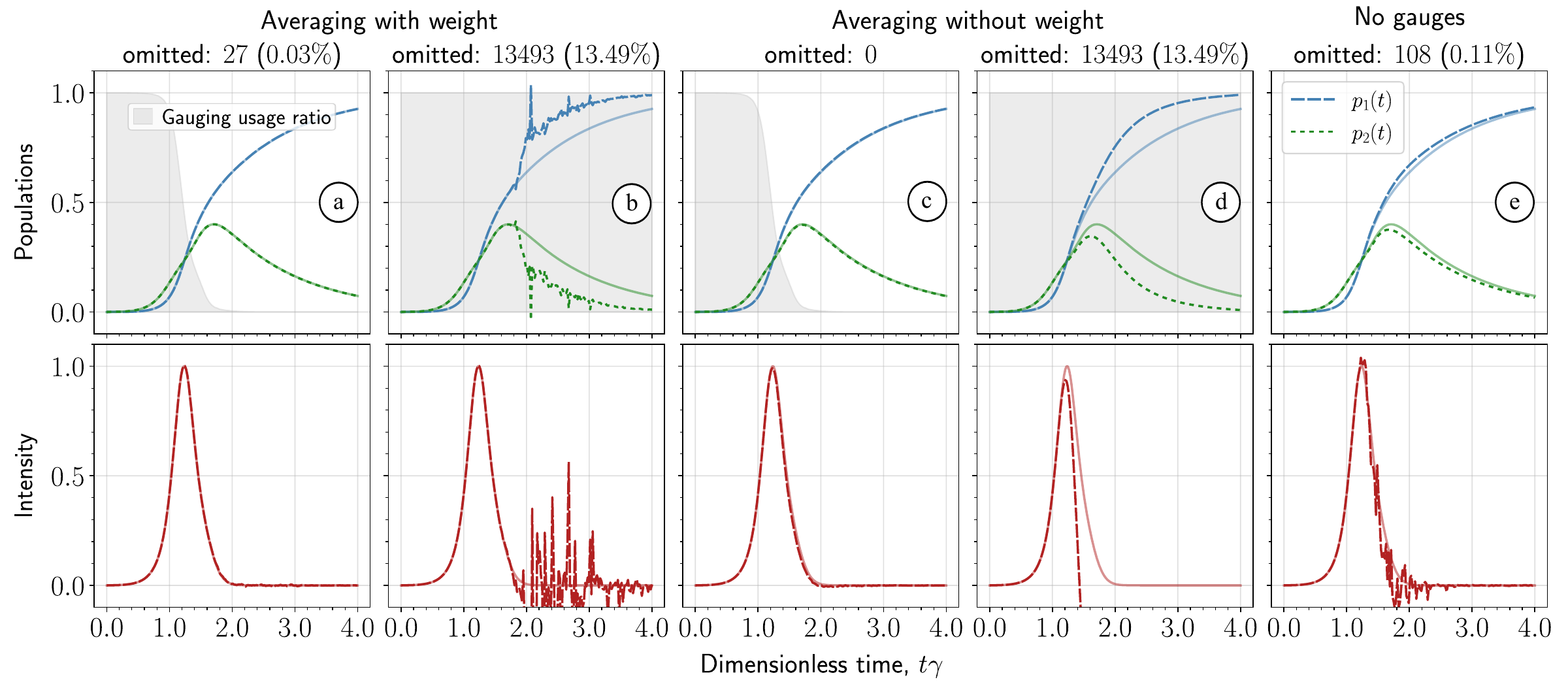}
    \caption{{ Numerical simulations of the population dynamics and intensity profiles of superfluorescence in 100 pumped two-level atoms studied in Sec.~\ref{Sec: quasi lambda}, comparing different gauging strategies. Here, we introduce stronger non-radiative decay with a rate of \(\Gamma=0.8\gamma\). The semi-transparent lines correspond to quantum expectation values, while the opaque lines represent stochastic averages.
(a) The drift gauge is applied only when population inversions are positive, with averaging performed with the weight function \(\Omega(t)\). The gray-filled area shows gauging frequency.
(b) Constant application of the drift gauge with the weight function. The results show strong intensity spikes and incorrect averages.
(c) Similar to panel (a), but averaging is done without the weight function. The expectation values are similar to (a) with a slight discrepancy in intensity around \(t\gamma = 2\).
(d) Constant application of the drift gauge without the weight function. The expectation values worsen compared to (b), with discrepancies appearing earlier.
(e) Ungauged simulations provided for comparison, demonstrating the necessity of gauging for accurate results.}}
    \label{fig: weight}
\end{figure*}

\subsection{Weight function}

{ After highlighting the importance of stochastic gauging, let us discuss some of its specifics.} {In all our numerical illustrations, we employ the drift gauging technique by modifying the deterministic terms only during the amplification of emission}, specifically when \(\mathrm{Re}(\rho_{ee} - \rho_{gg}) \geq 0\) for any excited state \(\ket{e}\) and ground state \(\ket{g}\). In exchange for this modification we introduce the weight function \(\Omega(t)\), which assigns a statistical weight to each trajectory. Additionally, we constantly apply the diffusion gauge that rescales noise terms. Although the drift gauge can, in principle, be applied constantly, this approach leads~\,to inaccurate simulations {~\,due~\,to~the exponential \newpage ~ \newpage ~ \newpage\noindent growth of the weight function}. However, when the gauging is applied carefully, following the guidance in Sec. \ref{sec: stochastic gauges}, our simulations yield correct results.

{Similarly to Sec. \ref{sec: stochastic gauges disc}, we consider an example of \(N=100\) pumped two-level atoms. Here, we introduce a stronger non-radiative decay, specifically, \(\Gamma = 0.8 \, \gamma\).}

In {Fig. \ref{fig: weight} (a)}, {the drift gauge is utilized only when there is an amplification of emission} and the statistical averages include the weight function \(\Omega(t)\). The gray-shaded area in the plot illustrates how frequently the drift gauge is applied. On average, the drift gauge is mostly applied until the mean intensity reaches its peak. After this peak, the probability of applying the gauge drops abruptly to zero.

In the opposite case, when the drift gauge is constantly applied, the obtained averages are incorrect, as demonstrated in Fig. \ref{fig: weight} {(b)}. More unstable trajectories are observed, and the expectation values exhibit many spikes. {This clearly shows that the drift gauge should be applied as rarely as possible and only when necessary.}

{When the drift gauge is used carefully and the system features sufficiently strong dissipation, the weight function \(\Omega(t)\) becomes less important. To demonstrate this, simulations in Fig. \ref{fig: weight} feature a higher dissipation rate compared to those in Fig. \ref{fig:gauge_on_and_off}. As shown in Fig. \ref{fig: weight} (c), if we apply the drift and diffusion gauge as suggested in Sec. \ref{sec: stochastic gauges} but without using the weight function, the expectation values remain almost the same as in Fig. \ref{fig: weight} (a). Only the intensity curve shows a small discrepancy around \(t\gamma = 2.0\). Additionally, since the weight coefficient does not enter expectation values, there are no unstable trajectories in the simulations shown in panel (c).}

{Notably, in the case when simulations are constantly gauged, omitting the weight function makes expectation values even worse, as suggested by comparing panel (d) with panel (b). Although the spikes disappear, discrepancies between the statistical and quantum-mechanical approaches appear earlier.}

{Remarkably, not using the weight function results in fewer discrepancies compared to the unnecessary application of the drift gauge. Since the weight function does not significantly impact the results, it is worth exploring whether gauging is necessary at all. Panel (e) displays the expectation values based on the ungauged equations. These expectation values considerably deviate from the exact solution, unlike cases in panels (a) and (c), proving that stochastic gauging is indeed necessary and improves the stability and accuracy of the simulations, even without the weight function.}

{It is important to note that omitting the weight function significantly changes the expectation values if the dissipation is not sufficiently strong. Specifically, if the simulations shown in Figs. \ref{fig:gauge_on_and_off} (b) and (g) had not included the weight function, the intensity curves would have featured a characteristic dip below zero, as observed in Fig. \ref{fig:gauge_on_and_off} (j). Therefore, we conclude that the weight coefficient can be safely disregarded when the system features strong dissipation processes, a condition typically encountered in experiments.}
\section{Conclusion}\label{conclusion}

The main achievement of this work is the development of a numerically efficient formalism, grounded in first principles, that reliably characterizes collective spontaneous emission for any number of multi-level emitters, especially in the presence of incoherent processes, within established bounds of applicability.

In this work, we focused on compact systems for benchmarking our methodology, as they can be solved exactly through quantum state decomposition. While our methodology generally yielded satisfactory results across a wide parameter range, we uncovered discrepancies when the system entered specific steady states. We showed that the presence of incoherent processes mitigated these issues.

Crucially, our formalism is characterized by equations that do not depend on the number of emitters, eliminating the numerical difficulties associated with traditional techniques based on quantum state decomposition.

Throughout the article, we have examined various scenarios and cases, from cooperative emission of instantly excited two-level atoms to complex multi-level systems, such as $V$- and $\Lambda$-type systems, showcasing the versatility and performance of our methodology.

{Our formalism has broad applicability for studying superfluorescence and superradiance across different spectral regions and emitter level structures. In particular, the approach offers an extension to previous methodologies \cite{Gross1982, PhysRevLett.42.224} and incorporates the preparation of the excited state manifold through pumping. While our numerical examples primarily focus on superfluorescence initiated by full inversion or incoherent pumping, the versatility of our proposed formalism enables the examination of systems characterized by initial macroscopic dipole moments and seeding fields resonant with atomic transitions. Additionally, the stochastic formalism can address superfluorescence in distributed media \cite{chuchurka2023stochastic, chuchurka2024hermitian}, allowing for the study of propagation effects.}

Our work offers valuable insights into the numerical challenges of simulating superfluorescence, the performance and limitations of our methodology, and its practical applicability in studying this fascinating quantum phenomenon.

\begin{acknowledgments}
    S.C. and V.S. acknowledge the financial support of Grant-No.~HIDSS-0002 DASHH (Data Science in Hamburg-Helmholtz Graduate School for the Structure of the Matter). {V.S. acknowledges the financial support of the Cluster of Excellence ``CUI: Advanced Imaging of Matter'' of the Deutsche Forschungsgemeinschaft (DFG) --- EXC 2056 --- project ID 390715994.}
\end{acknowledgments}

\bibliography{bib}

\appendix

\section{Stochastic gauges}\label{app: stochastic gauges}

In Refs.~\cite{2006'Deuar_stochastic-gauges,2005'Deuar_PhD}, the derivation of the stochastic gauge transformation was based on the freedom in decomposing the density matrix in terms of projectors constructed from coherent states. On one hand, the projectors are not defined uniquely; on the other hand, these projectors are analytical functions of their arguments, providing even more freedom. All of these observations indicate that the probabilistic interpretation of the density matrix is not unique. Since it is sampled using stochastic equations, the choice of these equations is also not unique. The most natural choice of stochastic equations does not necessarily lead to a stable numerical solution. Stochastic gauges offer the possibility of finding a more stable system of equations.

To apply stochastic gauges to our equations, we introduce stochastic freedom in a broader context. Let's start by investigating an arbitrary system of stochastic differential equations that yield a vector of stochastic processes, denoted as $\textbf{x}(t)$. The exact form of these equations and their origin may be disregarded in this appendix. Consider the following characteristic function:
\begin{equation}
\label{eq: char function definition}
    \chi(\bm{\lambda},t)=\Big\langle \exp{\big[\bm{\lambda}\cdot\textbf{x}(t)\big]}\Big\rangle.
\end{equation}
Its derivatives provide all the necessary information to calculate any expectation values of interest, namely:
\begin{equation*}
    \langle f[\textbf{x}(t)]\rangle = f\!\left[\frac{\partial}{\partial \bm{\lambda}}\right]\chi(\bm{\lambda},t)\Big|_{\bm{\lambda}=\textbf{0}}.
\end{equation*}
Consequently, $\chi(\bm{\lambda},t)$ is uniquely defined in the vicinity of $\bm{\lambda}=\textbf{0}$, since its derivatives at the point $\bm{\lambda}=\textbf{0}$ determine all observables.

Now, consider another system of equations that generates a different vector of stochastic processes, denoted as $\textbf{x}'(t)$, but yields exactly the same expectation values. The corresponding characteristic functions $\chi'(\bm{\lambda},t)$ must be identical to $\chi(\bm{\lambda},t)$:
\begin{equation}
\label{eq: condition for chi}
    \chi(\bm{\lambda},t)=\chi'(\bm{\lambda},t).
\end{equation} 

The explicit form of the stochastic trajectories $\textbf{x}(t)$ or $\textbf{x}'(t)$ is unknown, and only their stochastic differential equations are provided. Therefore, we cannot immediately construct the corresponding characteristic functions and compare them. We can only proceed in the spirit of mathematical induction. First, we ensure that the initial conditions for $\textbf{x}'(0)$ lead to the same characteristic function, satisfying Eq. (\ref{eq: condition for chi}) at $t=0$. Then, assuming that Eq. (\ref{eq: condition for chi}) holds for later times $t$, we guarantee that the temporal derivatives of the characteristic functions are preserved:
\begin{eqnarray}\label{eq: condition for chi 2}
    \frac{\partial}{\partial t} \chi(\bm{\lambda},t)&=&\frac{\partial}{\partial t}\chi'(\bm{\lambda},t)
\end{eqnarray}
The possibility of having multiple equivalent differential equations arises from the fact that the involved stochastic processes are, generally speaking, complex. In other words, the components of the vectors $\textbf{x}(t)$ and $\textbf{x}'(t)$ are, in reality, pairs of independent dynamic variables. However, the construction of expectation values does not involve these variables separately. In Eq.~(\ref{eq: char function definition}), the derivative with respect to $\lambda$ cannot extract only the real or imaginary part of $x_i(t)$. This property is the key source of stochastic freedom.

\section{Drift gauge}\label{app: drift gauge}

To provide an example of how the concept of stochastic gauge transformations from Appendix \ref{app: stochastic gauges} can be applied, we will derive the so-called drift gauge \cite{2005'Deuar_PhD} in the spirit of Girsanov's theorem \cite{liptser2001statistics}. In certain cases, stochastic differential equations take the form
\begin{equation*}
    \frac{d\textbf{x}(t)}{dt}=\textbf{A}(\textbf{x}(t),t)+\bm{\xi}(\textbf{x}(t),t)
\end{equation*}
where the drift terms $\textbf{A}(\textbf{x},t)$ can lead to diverging stochastic trajectories. Here, $\bm{\xi}(\textbf{x},t)$ represents Gaussian white noise terms with zero first moments and arbitrary second-order correlators. Unfortunately, neglecting diverging trajectories can result in incorrect expectation values. To tackle the numerical instability of divergent trajectories, one can opt for different stochastic equations with alternative drift terms. We will denote the new solution as $\textbf{x}'(t)$ and the alternative drift term as $\textbf{A}'(\textbf{x}',t)$. The new stochastic differential equations have the same initial conditions and read as follows:
\begin{equation*}
    \frac{d\textbf{x}'(t)}{dt}=\textbf{A}'(\textbf{x}',t)+\bm{\xi}(\textbf{x}'(t),t)
\end{equation*}
To compensate for this change, one can introduce a weight coefficient $\Omega(t) = e^{C_0(t)}$, which can be used to calculate expectation values based on the new stochastic variables:
\begin{equation*}
    \langle f(\textbf{x}(t))\rangle=\langle f(\textbf{x}'(t))\Omega(t)\rangle.
\end{equation*}
Consequently, the new characteristic function has the following form:
\begin{equation*}
    \chi'(\bm{\lambda},t)=\Big\langle \exp\!{\big(\bm{\lambda}\cdot\textbf{x}'(t)+C_0\!\left(t\right)\!\big)}\Big\rangle.
\end{equation*}
We can always formally write an equation of motion for $C_0(t)$:
\begin{equation*}
    \frac{dC_0(t)}{dt}=A_0(\textbf{x}'(t),t)+\xi_0(\textbf{x}'(t),t),
\end{equation*}
where $A_0$ is a new drift and $\xi_0$ is a new Gaussian white noise term. We assume that $\xi_0(\textbf{x},t)$ has a zero average and yet unknown correlation properties:
\begin{equation*}
\begin{gathered}
    \langle \xi_0(\textbf{x},t)\xi_0(\textbf{x},t')\rangle=\sigma_{0}(\textbf{x},t)\delta(t-t'), \\
    \langle \bm{\xi}(\textbf{x},t)\xi_0(\textbf{x},t')\rangle=\bm{\sigma}(\textbf{x},t)\delta(t-t').
\end{gathered}
\end{equation*}
The main goal is to find the drift $A_0$ for the weight coefficient and the correlation properties $\sigma_{0}$ and $\bm{\sigma}$ that compensate for the change in drift terms $\Delta \textbf{A} = \textbf{A}' - \textbf{A}$ at the level of the characteristic function. Following Appendix~\ref{app: stochastic gauges}, we proceed in the spirit of mathematical induction and assume that $\chi'\!\left(\bm{\lambda},t\right) = \chi\!\left(\bm{\lambda},t\right)$ is satisfied for a certain $t$. Let's check if the same holds for the derivatives: 
\begin{multline*}
    \frac{\partial}{\partial t}\left[\chi'\!\left(\bm{\lambda},t\right)-\chi\!\left(\bm{\lambda},t\right)\right]\\=\Bigg(\bm{\lambda}\cdot \left[ \Delta\textbf{A}\left(\frac{\partial}{\partial \bm{\lambda}},t\right)+\bm{\sigma}\left(\frac{\partial}{\partial \bm{\lambda}},t\right)\right]\\A_0\left(\frac{\partial}{\partial \bm{\lambda}},t\right)+\frac{1}{2}\sigma_{0}\left(\frac{\partial}{\partial \bm{\lambda}},t\right)\Bigg)\chi\!\left(\bm{\lambda},t\right).
\end{multline*}
In the derivation of this expression, we have used Itô's lemma from Eq. (\ref{app: ito}). To make the right-hand side equal to zero for any $\bm{\lambda}$, we have to choose the following correlation properties for the noise terms:
\begin{equation*}
\begin{gathered}
    \langle \xi_0(\textbf{x},t)\xi_0(\textbf{x},t')\rangle=-2A_0(\textbf{x},t)\delta(t-t'), \\
    \langle \bm{\xi}(\textbf{x},t)\xi_0(\textbf{x},t')\rangle=-\Delta\textbf{A}(\textbf{x},t)\delta(t-t').
\end{gathered}
\end{equation*}
This constitutes the essence of the drift gauge. Notably, our derivations are not based on the properties of projectors used to decompose the density matrix; our result is applicable to any system of stochastic trajectories, including the modified Bloch equations for the variables ${\rho}_{pq}(t)$.

\section{Diffusion gauge}\label{app:Diffusion gauge}
In this section, we provide the expressions for $\eta_\alpha(t)$ and $\theta_\alpha(t)$ used in our numerical simulations. These expressions are derived by substituting the decomposition from Eq.~(\ref{eq: decomposition of the new elementary noise terms}) into the expression in Eq.~(\ref{eq: function to be minimized}). Subsequently, we minimize this expression with respect to $\eta_\alpha(t)$ and $\theta_\alpha(t)$, aiming to mitigate the amplification of non-Hermitian components. The resulting functions, $\eta_\alpha(t)$ and $\theta_\alpha(t)$, take the following form:
\begin{widetext}
\begin{align*}
    \eta_{\beta}^4(t) 
    = 
    \dfrac{\sum_{\alpha} \left( \left|P_{\alpha\beta}^{(ee)}(t) - P_\alpha^{(+)}(t) P_{\beta}^{(-)}(t) \right|^2 + \left| P_\alpha^{(-)}(t) P_{\beta}^{(-)}(t) \right|^2 \right) }{\sum_{\alpha} \left(  \left|P_{\beta\alpha}^{(ee)}(t) - P_{\alpha\beta}^{(gg)}(t) \right|^2 \right)},
\end{align*}
\begin{align*}
    \theta_{\beta}^4(t)  = \dfrac{\sum_{\alpha} \left(  \left|P_{\beta\alpha}^{(ee)}(t) - P_{\beta}^{(+)}(t) P_\alpha^{(-)}(t) \right|^2 + \left| P_\alpha^{(+)}(t) P_{\beta}^{(+)}(t) \right|^2\right) }{\sum_{\alpha} \left(  \left|P_{\alpha\beta}^{(ee)}(t) - P_{\beta\alpha}^{(gg)}(t) \right|^2 \right)},
\end{align*}
\end{widetext}
where for simplicity we introduced the following tensors:
\begin{align*}
& P^{(gg)}_{\alpha\beta}(t) = \sum_{e,g,g'} d_{eg,\alpha} \, \rho_{gg'}(t) \, d_{g'e, \beta},
\\
& P^{(ee)}_{\alpha\beta}(t) = \sum_{e,e',g} d_{ge,\alpha} \, \rho_{ee'}(t) \, d_{e'g, \beta}.
\end{align*}
and the following vectors that are stochastic counterparts of the polarization fields:
\begin{align*}
& \mathbf{P}^{(+)}(t) = \sum_{e,g} \mathbf{d}_{ge} \, \rho_{eg}(t),\\
& \mathbf{P}^{(-)}(t) = \sum_{e,g} \mathbf{d}_{eg} \, \rho_{ge}(t).
\end{align*}

\end{document}